\documentclass[11pt]{amsart}

\usepackage{amsmath}
\usepackage{amsfonts}
\usepackage{amssymb}
\usepackage{amscd}
\usepackage{amsthm}
\usepackage{framed}
\usepackage{fullpage}
\usepackage{graphicx}
\usepackage{latexsym}
\usepackage[numbers]{natbib}  
\usepackage{multirow}
\usepackage{tikz}
\usetikzlibrary{positioning}
\usetikzlibrary{arrows}

\allowdisplaybreaks[4]

\usepackage{hyperref}
\hypersetup{colorlinks,linkcolor={red},citecolor={blue},urlcolor={blue}}

\theoremstyle{definition}

\numberwithin{equation}{section}

\newcommand{\CC}{\mathbb C}

\newcommand{\cW}{\mathcal W}

\newcommand{\ZZ}{\mathbb Z}

\def\dim{\operatorname{dim}}

\newcommand{\Eh}{E_{7+1/2}}
\newcommand{\Dh}{D_{6+1/2}}

\newcommand{\Ah}{A_{5+1/2}}

\newcommand{\mg}{\mathfrak{g}}

\def\hv{{h^\vee}}
\newcommand{\T}{\mathsf{T}}
 
\usepackage{multirow}

\newcommand{\cF}{\mathcal F}
\newcommand{\cB}{\mathcal B}

\newcommand{\TF}{\mathsf{T}^{\rm F}}

\begin{document}

\title[Minimal $\mathcal{W}$-algebras with non-admissible levels and intermediate Lie algebras]{Minimal $\mathcal{W}$-algebras with non-admissible levels and intermediate Lie algebras}

\author{Kaiwen Sun}

\address{Department of Physics and Astronomy, Uppsala University, 75120, Uppsala, Sweden\newline
\indent Department of Mathematics, Uppsala University, 75106, Uppsala, Sweden}

\email{kaiwen.sun@physics.uu.se}

\subjclass[2020]{}

\date{\today}


\begin{abstract} 
In \cite{Kawasetsu:2018irs}, Kawasetsu proved that the simple $\mathcal{W}$-algebra associated with a minimal nilpotent element $\mathcal{W}_{k}(\mathfrak{g},f_\theta)$ is rational and $C_2$-cofinite for $\mathfrak{g}=D_4,E_6,E_7,E_8$ with non-admissible level $k=-h^\vee/6$. In this paper, we study $\mathcal{W}_{k}(\mathfrak{g},f_\theta)$ algebra for $\mathfrak{g}=E_6,E_7,E_8$ with non-admissible level $k=-h^\vee/6+1$. We determine all irreducible (Ramond twisted)  modules, compute their characters and find coset constructions and Hecke operator interpretations. These $\mathcal{W}$-algebras are closely related to intermediate Lie algebras and intermediate vertex subalgebras.

\end{abstract}

\maketitle
%\begin{small}\tableofcontents\end{small}

\section{Introduction}

$\cW$-algebras are first introduced by Zamolodchikov  as higher spin extensions of Virasoro algebras \cite{zamolodchikov1985infinite}. In general, the $\cW$-algebras $\cW^k(\mg,f)$ are defined associated to a finite dimensional simple Lie algebra $\mg$, a nilpotent element $f\in \mg$ and a complex number $k$ called level. They are constructed by the quantized Drinfel'd-Sokolov reduction from affine vertex operator algebras \cite{Feigin:1990pn,Kac:2003jh}. 
A level $k$ is called $\textit{admissible}$ if 
\begin{align}
    h^\vee+k=\frac{p}{q},\quad p,q\in \ZZ_{>0},\ (p,q)=1\ \text{and}\ p\ge \left\{  
             \begin{array}{ll}  
             \!h^\vee  &\ (r^\vee,q)=1, \\  
             \! h  &  \   (r^\vee,q)= r^\vee,
             \end{array}  
\right.
\end{align}
where $h$ is the Coxeter number, $h^\vee$ the dual Coxeter number and $r^\vee$  the lacing number of $\mg$.  
The rationality and $C_2$-cofiniteness of $\cW$-algebras with admissible levels have been conjectured and proved for many types of $(\mg,k,f)$  \cite{Frenkel:1992ju,kac2008rationality,arakawa2015associated,arakawa2015rationality,arakawa2023rationality}. However, the $\cW$-algebras with non-admissible levels are more difficult to study and rational examples are much less known \cite{Kawasetsu:2018irs,Creutzig:2021dda,fasquel2022geometry,Milas:2023cwx}.

Denote the simple quotient VOA of $\cW^k(\mg,f)$ as $\cW_k(\mg,f)$. In this paper, we are interested in the \textit{minimal $\cW$-algebras} $\cW_k(\mg,f)$ where $\mathfrak{g}=E_{6,7,8}$, $f$ is the minimal nilpotent element $f_\theta$ and $k\ge -h^\vee/6-1$ is a negative integer. For $k=-h^\vee/6-1$, $\cW_k(\mg,f_\theta)\cong \CC$ is trivial. For $k=-h^\vee/6$, Kawasetsu proved that $\cW_k(\mg,f_\theta)$ is $\ZZ_2$-rational and $C_2$-cofinite \cite{Kawasetsu:2018irs}. For $-h^\vee/6+1\le k\le -1$, Arakawa and Moreau proved that $\cW_k(\mg,f_\theta)$ is $C_2$-cofinite \cite{AM}. These results suggest that the minimal $\cW$-algebras under consideration have remarkably nice properties even though they are with non-admissible levels. 

The aim of the present paper is to study these minimal $\cW_k(\mg,f_\theta)$ algebras with non-admissible level  $-h^\vee/6\le k\le -1$ by physical techniques in 2d CFTs such as coset construction, Hecke operator, simple current, non-diagonal modular invariant, bosonization and fermionization and so on. In particular, for $k=-h^\vee/6+1$, we are able to work out all irreducible modules of the $\cW$-algebras. The main results can be summarized as follows. 
For $\mathfrak{g}=E_{6,7,8}$, let $\mathfrak{g}_{*}\times A_1$ be the maximal subalgebra of $\mathfrak{g}$, then $L_2( \mathfrak{g}_{*})$ is the maximal affine sub VOA of $\cW_{- h^\vee/{6}+1}(\mathfrak{g},f_\theta)$. Denote $m=24/(h^\vee-6)$, we find the following coset construction holds:
\begin{align}\label{cosetWE}
    \frac{\cW_{- h^\vee/{6}+1}(\mathfrak{g},f_\theta)}{L_2( \mathfrak{g}_{*})}   =\cB\mathrm{Vir}^{N=1}_{3m+10,m+4}.
\end{align}
Here $\cB$ is the bosonization of $N=1$ SVOAs. 
More precisely, we have the following results. 
\\[+2.5mm]
\textbf{Coset Construction A.} \textit{
The simple $\cW$-algebras 
$\mathcal{W}_{-1}(E_6,f_{\theta})$, $\mathcal{W}_{-2}(E_7,f_{\theta})$ and $\mathcal{W}_{-4}(E_8,f_{\theta})$ are $\ZZ_2$-rational and $C_2$-cofinite with $27,13$ and $6$ irreducible modules respectively. 
    There exist coset constructions:
    \begin{align} \label{cosetWE6}
    \frac{\mathcal{W}_{-1}(E_6,f_{\theta})}{L_2(A_5)} & = (\cB\mathrm{Vir}^{N=1}_{ 22,8})^{\rm e},\\ \label{cosetWE7}
        \frac{\mathcal{W}_{-2}(E_7,f_{\theta})}{L_2(D_6)} & = \cB\mathrm{Vir}^{N=1}_{16 ,6},\\ \label{cosetWE8}
        \frac{\mathcal{W}_{-4}(E_8,f_{\theta})}{L_2(E_7)} & = \cB\mathrm{Vir}^{N=1}_{ 13,5} .
    \end{align}
The three VOAs on the right hand sides are the bosonizations of $N=1$ supersymmetric Virasoro minimal models and extensions with $49 ,58$ and $36$ irreducible modules respectively. Denote them as $M_6,M_7$ and $M_8$, then the following isomorphisms hold:
\begin{align}  \nonumber
  \ \ \ \ \ {\mathcal{W}_{-1}(E_6,f_{\theta})}  \cong &\, L_2(A_5)\otimes M_6(0)\oplus L_2(A_5, 2w_3)\otimes M_6 \bigg(\frac{ 3}{ 2}\bigg)\oplus L_2(A_5,w_1+w_5)\otimes M_6\bigg(\frac94 \bigg)\\ \label{isoE6}
    &\oplus L_2(A_5, w_2+w_4)\otimes M_6 \bigg(\frac{ 11}{ 4}\bigg)\oplus L_2(A_5, w_3)\otimes M_6 \bigg(\frac{  27}{ 32}\bigg)\\ \nonumber
    & \oplus L_2(A_5, w_1+w_2)\otimes M_6 \bigg(\frac{  143}{ 32}\bigg)\oplus L_2(A_5, w_4+w_5)\otimes M_6 \bigg(\frac{  143}{ 32}\bigg),\\  \nonumber
       {\mathcal{W}_{-2}(E_7,f_{\theta})}  \cong &\, L_2(D_6)\otimes M_7(0)\oplus L_2(D_6, 2w_6)\otimes M_7 \bigg(\frac{ 3}{ 2}\bigg)\oplus  L_2(D_6,2w_1)\otimes M_7(7)\\  \label{isoE7}
 &\oplus  L_2(D_6,2w_5)\otimes M_7\bigg(\frac{15 }{2   }\bigg) \oplus  L_2(D_6,w_2)\otimes M_7\bigg(\frac{13 }{6   }\bigg)\oplus  L_2(D_6,w_4)\otimes M_7\bigg(\frac{8 }{3  }\bigg)\\ \nonumber
 &\oplus  L_2(D_6,w_6)\otimes M_7\bigg(\frac{ 13}{16   }\bigg)\oplus  L_2(D_6,w_1+w_5)\otimes M_7\bigg(\frac{69 }{  16 }\bigg),\\   \nonumber
         \cW_{-4}(E_8,f_\theta)  \cong &\, L_2(E_7)\otimes M_8(0)\oplus  L_2(E_7,2w_6)\otimes M_8\bigg(\frac32\bigg)\oplus  L_2(E_7,w_1)\otimes M_8\bigg(\frac{21}{10}\bigg) \\   \label{isoE8}
 &\oplus  L_2(E_7,w_5)\otimes M_8\bigg(\frac{13}{5}\bigg) \oplus L_2(E_7,w_6)\otimes M_8\bigg(\frac{63}{80}\bigg)\oplus  L_2(E_7,w_7)\otimes M_8\bigg(\frac{67}{16}\bigg).
    \end{align}
    The $\cW$-algebras have $\ZZ_2$ automorphism group $\{1,\sigma\}$  such that in the decomposition to $L_2(\mg_*)\otimes M$ modules, the last three modules in \eqref{isoE6}, the last two in \eqref{isoE7} and the last two in \eqref{isoE8} are odd for $\sigma$, while the rest are even.
    }

To establish the above isomorphisms, our strategy is to first use Hecke operator \cite{Harvey:2018rdc} or fermionic Hecke operator \cite{Lee:2022yic} to produce the vector-valued modular form composed of the characters of Ramond-twisted irreducible modules of the $\cW$-algebra, then we find the coset construction for the Ramond-twisted modules by verifying the character relations. Finally, we recover all irreducible modules from the Ramond-twisted ones by a $\ZZ_2$ automorphism of $L_2(\mg_*)$ irreducible modules induced by a simple current of $L_2(\mg_*)$ with conformal dimension $3/2$, which is the supersymmetry generator in the fermionization of $L_2(\mg_*)$. The proof should follow the approach of \cite{Kawasetsu:2018irs} for the cases with $k=-h^\vee/6$.

The vector-valued modular forms that describe the characters of Ramond-twisted irreducible modules of $\mathcal{W}_{-4}(E_8,f_{\theta})$ and $\mathcal{W}_{-2}(E_7,f_{\theta})$ have been realized by Hecke operator $\T_{19}$ on $\mathrm{Vir}_{13,2}$ \cite{Duan:2022ltz} and by fermionic Hecke operator $\TF_{11}$ on $\mathrm{Vir}_{16,2}^{N=1}$ \cite{Lee:2024fxa} respectively. In this paper, we newly find the following useful Hecke relation.
\\[+3mm]
\noindent\textbf{Hecke operator.} \textit{The characters of Ramond-twisted irreducible modules of 
$\mathcal{W}_{-1}(E_6,f_{\theta})$ are realized by Hecke operator $\T_7$ on an extension of $\cB\mathrm{Vir}^{N=1}_{132 ,2}$.}
\\[-0.5mm]

The minimal $\cW_k(\mg,f_\theta)$ algebras with non-admissible levels under consideration are closely related to the \textit{intermediate vertex subalgebras} introduced by Kawasetsu in \cite{Kawasetsu}, which are often called \textit{intermediate VOAs} in 2d CFT literature. They can be regarded as the generalization of affine VOA at positive integer level for simple Lie algebra to \textit{intermediate Lie algebra}. 
The first example of such exotic subjects appeared in the 
Mathur--Mukhi--Sen classification on the CFT-type solutions of the second order holomorphic modular linear differential equation (MLDE) \cite{Mathur:1988na}. A special solution with central charge $38/5$ was found and expected to describe a WZW model at level one for a non-reductive Lie algebra between $E_7$ and $E_8$ \cite{Mathur:1988na}. This non-reductive Lie algebra has dual Coxeter number $24$ and dimension $190$ and was later constructed by Landsberg and Manivel using sextonions \cite{LM}. It is called the intermediate Lie algebra $\Eh$, which fills a hole between $E_7$ and $E_8$ in the Cvitanovi\'c--Deligne exceptional series \cite{Cvitanovic:2008zz,Deligne}.   
The exotic WZW model found in \cite{Mathur:1988na} can then be formally denoted as $L_1(\Eh)$. Precisely speaking, this is an intermediate vertex subalgebra of $E_8$ lattice VOA \cite{Kawasetsu}. A more interesting understanding is that it characterizes the Ramond-twisted irreducible modules of minimal $\cW$-algebra $W_{-5}(E_8,f_\theta)$ \cite{Kawasetsu:2018irs}. 

The correspondence between $L_1(\Eh)$ and $W_{-5}(E_8,f_\theta)$ can be generalized to all $W_{-h^\vee/6}(\mg,f_\theta)$ for $\mg$ belonging to the Cvitanovi\'c--Deligne exceptional series. Accordingly $L_1(\Eh)$ is generalized to $L_1(\mg_I)$ where $\mg_I$ belongs to a series of non-reductive Lie algebras called the \textit{intermediate exceptional series} satisfying $\mg_*\subset \mg_I\subset \mg$ \cite{Lee:2024fxa} and $L_1(\mg_I)$ is an intermediate vertex subalgebra of $L_1(\mg)$. Then the characters of Ramond twisted irreducible modules of $W_{-h^\vee/6}(\mg,f_\theta)$ algebra \cite{Kawasetsu:2018irs} coincide with the characters of irreducible modules of $L_1(\mg_I)$. These characters also satisfy a one-parameter family of 4th order holomorphic MLDE \cite{Kawasetsu:2018tzs}.

It is natural to consider the intermediate vertex subalgebra $L_k( \mathfrak{g}_{I})$ with higher levels, which are expected to be related to minimal $\cW$-algebra with higher levels. Indeed, in this paper we find for $\mg=E_{6,7,8}$, the characters of Ramond twisted irreducible modules of $W_{-h^\vee/6+1}(\mg,f_\theta)$ coincide with the characters of $L_2(\mg_I)$. This relies on the observation that the coset construction \eqref{cosetWE} can also be written as follows
\begin{align}\label{cosetWEeff}
    \frac{L_2( \mathfrak{g}_{I})}{L_2( \mathfrak{g}_{*})}   =(\cB\mathrm{Vir}^{N=1}_{3m+10,m+4})_{\rm eff}.
\end{align}
The $L_2( \mathfrak{g}_{I})$ is an intermediate vertex subalgebra of $L_2(\mg)$, and is somewhat easier to study than $W_{-h^\vee/6+1}(\mg,f_\theta)$ as they are $\ZZ$-graded rather than $\ZZ/2$-graded. Some efforts have been made recently to find the conformal dimensions, characters and coset constructions of $L_k( \mathfrak{g}_{I})$ for $k\ge 2$ \cite{Lee:2023owa,Lee:2024fxa}. Below an upper bound of level $k$, the intermediate vertex subalgebra $L_k(\mg_I)$ has central charge $k\dim(\mg_I) /(h^\vee+k)$ just like usual affine VOAs. For the purpose of present paper, we are interested in $\mg_I=\Ah,\Dh,\Eh$ which have dimensions $56,99,190$ and dual Coxeter numbers $9,14,24$ respectively.  The former two are defined by intermediate subalgebra relations $A_5\subset \Ah\subset E_6$ and $D_6\subset \Dh\subset E_7$.

An interesting coset construction $L_2(E_8)/L_2(\Eh)=\mathrm{Vir}_{13 ,4}^{\rm eff}$ was found in \cite{Lee:2023owa}. In this paper we further find this type of coset construction for $L_2(\Dh)$ and $L_2(\Ah)$. Such cosets can be written uniformly for $\mg=E_{6,7,8}$ as follows.
\\[+2.5mm]
\textbf{Coset Construction B.} \textit{The characters of Ramond-twisted irreducible modules of $\cW_{-h^\vee/6+1}(\mathfrak{g},f_\theta)$ coincide with the characters of irreducible modules of 
$L_2(\mathfrak{g}_{I})$, which is an intermediate vertex subalgebra of $L_2(\mathfrak{g})$. The following coset construction holds:
    \begin{align}
    \frac{L_2(\mathfrak{g}) }{ L_2(\mathfrak{g}_{I})}=\mathrm{Vir}_{3m+10 ,m+3}^{\rm eff}.
\end{align}
}
\\[+2.5mm]
More precisely, for $\mg=E_7$, the coset gives $\mathrm{Vir}_{16 ,5}^{\rm eff}$, while for $\mg=E_6$, it gives an extension that is the type-$(A_6,D_{12})$ non-diagonal modular invariant of $\mathrm{Vir}_{22 ,7}^{\rm eff}$. 
As comparison, 
$L_1(\mathfrak{g}_{I})$ is intermediate vertex subalgebra of $L_1(\mathfrak{g})$ and there exist coset constructions
    \begin{align}
    \frac{L_1(\mathfrak{g}) }{ L_1(\mathfrak{g}_{I})}=\mathrm{Vir}_{5 ,2}^{\rm eff}.
\end{align}

The paper is organized as follows. In Section \ref{sec:rev}, we review the various properties of $\cW_{-h^\vee/6}(\mathfrak{g},f_\theta)$ algebras  studied in \cite{Kawasetsu:2018irs} for $\mg$ belonging to the Cvitanovi\'c--Deligne exceptional series and their connection with intermediate vertex subalgebras $L_1(\mg_I)$. In Section \ref{sec:E8}, \ref{sec:E7}, \ref{sec:E6}, we discuss $\cW_{-h^\vee/6+1}(\mathfrak{g},f_\theta)$ for $\mg=E_8,E_7,E_6$ respectively, including coset constructions, irreducible modules, Ramond twisted irreducible modules and characters. In Section \ref{sec:E8high}, we make some observations for $\cW_{k}(E_8,f_\theta)$ with $k=-3,-2$, and point out a series of Hecke relations between some subregular exceptional $\cW$-algebras with admissible levels and the minimal $\cW$-algebras with non-admissible levels.

Notation: We use $\cF$ for fermionization and $\cB$ for bosonization. We use $\phi$ to denote the bosonic primaries from $\mathrm{NS}+\widetilde{\mathrm{NS}}$ sector, i.e., the even part of Neveu-Schwarz modules, $\varphi$ from $\mathrm{NS}-\widetilde{\mathrm{NS}}$ sector, i.e., the odd part of Neveu-Schwarz modules, and $ \psi$ from R sector, i.e., the Ramond-twisted modules. The Virasoro minimal model $\mathrm{Vir}_{p,q}$ and the supersymmetric version $\mathrm{Vir}_{p,q}^{N=1}$ are often denoted as $M(p,q)$ and $SM(p,q)$ in physics literature. The VOA extension denoted as $( )^{\rm e}$ is often called non-diagonal modular invariant in physics literature. We adopt the Dynkin diagrams and representation names of simple Lie algebras from \href{https://lieart.hepforge.org/}{LieART} package of Mathematica as in previous works \cite{Lee:2023owa,Lee:2024fxa}. The simple currents and fusion rules of $L_k(\mg)$ are computed from \href{https://www.nikhef.nl/~t58/Site/Kac.html}{Kac}, while the characters of $L_k(\mg)$ are computed from \href{https://www.sagemath.org/}{SageMath}.

\section{Review of $\cW_{k}(\mathfrak{g},f_\theta)$ at level $k=-h^\vee/6$}\label{sec:rev}

The $\cW_k(\mathfrak{g},f_\theta)$ algebra associated with a minimal nilpotent element has central charge \cite{Kac:2003jh}
\begin{equation}
    c_{\cW}=\frac{k\dim\mathfrak{g}}{k+\hv}-6k+\hv-4.
\end{equation}
Let $\mg$ be a simple Lie algebra in the Cvitanovi\'c--Deligne exceptional series \cite{Cvitanovic:2008zz,Deligne}
$$
A_1\subset A_2 \subset G_2 \subset D_4 \subset F_4 \subset E_6 \subset E_7 \subset E_8.
$$
The $\cW_k(\mathfrak{g},f_\theta)$ algebras with level $k=-h^\vee/6$ have been studied by Kawasetsu in \cite{Kawasetsu:2018irs}. 
For $\mathfrak{g}=A_1,A_2,G_2,F_4$, the level $k=-h^\vee/6$ is admissible, while for $\mathfrak{g}=D_4,E_6,E_7,E_8$, the level is non-admissible. 
Except $\cW_{-1/3}(A_1,f_\theta)$ is isomorphic to Virasoro minimal model $\mathrm{Vir}_{5,3}$, the rest cases are rather non-trivial. Remarkably, Kawasetsu proved that for $\mathfrak{g}=D_4,E_6,E_7,E_8$, the $\cW_{-h^\vee/6}(\mathfrak{g},f_\theta)$ is $\ZZ_2$-rational and $C_2$-cofinite \cite{Kawasetsu:2018irs}, which contribute precious examples of rational $\cW$-algebras with non-admissible levels. It was also proved in \cite{Kawasetsu:2018irs} that the characters of the Ramond-twisted irreducible modules of $\cW_{-h^\vee/6}(\mathfrak{g},f_\theta)$ algebra forms a vector-valued modular form on $SL(2,\ZZ)$. In the following we review Kawasetsu's results in a way that leads to similar yet more intricate construction for $\cW_{-h^\vee/6+1}(\mathfrak{g},f_\theta)$ algebras.

Denote $L(c,h)$ as the irreducible highest weight module of Virasoro algebra of central charge $c$ with highest weight $h$. Recall $\mathrm{Vir}_{5,3}$ has central charge $-3/5$ and four irreducible highest weight modules with highest weights $-1/20,0,1/5,3/4$. One main result of \cite{Kawasetsu:2018irs} is that for $\mg=E_{6,7,8},$ the following isomorphism holds
\begin{align}\label{Wiso1}
   \cW_{-h^\vee/6}(\mg,f_\theta) \cong L_1 (\mg_*)\otimes L(-3/5,0)\oplus L_1 (\mg_*,w_*)\otimes L(-3/5,3/4).
\end{align}
Here the $L_1 (\mg_*,w_*)$ is a $\ZZ_2$-simple current of $L_1 (\mg_*)$ with conformal dimension $3/4$. The $L_1 (\mg_*)\otimes L(-3/5,0)$ is even while $L_1 (\mg_*,w_*)\otimes L(-3/5,3/4)$ is odd for the $\ZZ_2$ automorphism group of $\cW_{-h^\vee/6}(\mg,f_\theta)$ \cite{Kawasetsu:2018irs}. For $\mg=D_4$, similar isomorphism holds except $\mg_*=A_1^3$ is semisimple. By physics notation, this can be written as the coset 
\begin{align}\label{coset1}
   \frac{\cW_{-h^\vee/6}(\mg,f_\theta) }{ L_1 (\mg_*)}=\mathrm{Vir}_{5,3}.
\end{align}

All irreducible modules and Ramond-twisted irreducible modules of $\cW_{-h^\vee/6}(\mg,f_\theta)$ can then be determined.  
For example, for $\cW_{-5}(E_8,f_\theta)$, the two Ramond twisted irreducible modules are \cite{Kawasetsu:2018irs}
\begin{align}\nonumber
    R_0& =L_1 (E_7)\otimes L(-3/5,-1/20)\oplus L_1 (E_7,w_6)\otimes L(-3/5,1/5),\\\nonumber
    R_1& =L_1 (E_7)\otimes L(-3/5,3/4)\oplus L_1 (E_7,w_6)\otimes L(-3/5,0).
\end{align}
The characters of $R_0,R_1$ are exactly the exotic $c=38/5$ solution of 2nd order MLDE found by Mathur--Mukhi--Sen in \cite{Mathur:1988na}, i.e., the characters of $L_1(\Eh)$. They are also realized as a $\T_{19}$ Hecke image of the characters of $\mathrm{Vir}_{5,2}$ in \cite{Harvey:2018rdc}. On the other hand, the two irreducible modules of $\cW_{-5}(E_8,f_\theta)$ are \cite{Kawasetsu:2018irs}
\begin{align}\nonumber
    N_0& =L_1 (E_7)\otimes L(-3/5,0)\oplus L_1 (E_7,w_6)\otimes L(-3/5,3/4),\\\nonumber
    N_1& =L_1 (E_7)\otimes L(-3/5,1/5)\oplus L_1 (E_7,w_6)\otimes L(-3/5,-1/20).
\end{align} 
The $\ZZ_2$ automorphism group $\{1,\sigma\}$ of $\cW_{-5}(E_8,f_\theta)\cong N_0$ acts as $\sigma(b)=b$ and $\sigma(f)=-f$ for any $b\in L_1 (E_7)\otimes L(-3/5,0)$ and $f\in L_1 (E_7,w_6)\otimes L(-3/5,3/4)$ \cite{Kawasetsu:2018irs}, which means in the decomposition to $L_1(E_7)\otimes L(-3/5,0)$ modules, the first module in $N_0$ is even while the second is odd. 
Notably, the $E_7$ irreducible module $R_{w_6}=\bf 56$ is a fermionic representation and $w_6$ is a fermionic weight in the sense of \cite{Lee:2023owa,Lee:2024fxa}. This suggests the concepts of fermionic representation and fermionic weight in \cite{Lee:2023owa,Lee:2024fxa} are directly linked to the $\ZZ_2$ automorphism of the $\cW_{k}(\mg,f_\theta)$ algebras.

Notice that $L_1 (E_7,w_6)$ is the $\ZZ_2$-simple current of $L_1 (E_7)$, which induces an exchange between $L_1 (E_7)$ and $L_1 (E_7,w_6)$. The irreducible modules $N_0,N_1$ of $\cW_{-5}(E_8,f_\theta)$ are recovered from the Ramond-twisted ones $R_0,R_1$ by such exchange. This nice relation between irreducible modules and Ramond-twisted ones persists for  other more complicated $\cW$-algebras.   

For $\cW_{-3}(E_7,f_\theta)$, the four Ramond twisted irreducible modules are
\begin{align}\nonumber
   R_0& =L_1 (D_6)\otimes L(-3/5,-1/20)\oplus L_1 (D_6,w_6)\otimes L(-3/5,1/5),\\\nonumber
    R_1& =L_1 (D_6)\otimes L(-3/5,3/4)\oplus L_1 (D_6,w_6)\otimes L(-3/5,0),\\\nonumber
    R_3& =L_1 (D_6,w_1)\otimes L(-3/5,0)\oplus L_1 (D_6,w_5)\otimes L(-3/5,3/4),\\\nonumber
    R_4& =L_1 (D_6,w_1)\otimes L(-3/5,1/5)\oplus L_1 (D_6,w_5)\otimes L(-3/5,-1/20).
\end{align}
The characters of $R_0,R_1,R_2,R_3$ coincide with the characters of $L_1(\Dh)$ with conformal dimension $0,4/5,11/20,3/4$ respectively \cite{Lee:2024fxa}. As vector valued modular form, they can be realized by $\T_{11}$ Hecke image of $(\mathrm{Vir}_{5,3})_{\rm eff}$ \cite{Lee:2024fxa}. The above module decompositions are consistent with the character relations in \cite[Equations (4.87)--(4.90)]{Lee:2024fxa}.

To recover the irreducible modules of $\cW_{-3}(E_7,f_\theta)$, consider the simple current of $L_1(D_6)$. It is well-known that 
$L_1(D_6)$ has two $\ZZ_2$-simple currents $L_1 (D_6,w_5)$ and $L_1 (D_6,w_6)$ which are conjugate to each other. It is conventional to choose $L_1 (D_6,w_6)$ which corresponds to the conjugate spinor representation   $\bf\overline{32}$ of $D_6$. This simple current induces a $\ZZ_2$ isomorphism for the irreducible $L_1(D_6)$ modules that is the exchange between $L_1(D_6)$ and $L_1 (D_6,w_6)$, and the exchange between $L_1 (D_6,w_1)$ and $L_1 (D_6,w_5)$. 
The four irreducible modules of $\cW_{-3}(E_7,f_\theta)$ can then be recovered from the Ramond-twisted ones by such $\ZZ_2$ isomorphism. For example, $R_1$ is mapped to $N_0\cong \cW_{-3}(E_7,f_\theta)$ in \eqref{Wiso1}. Besides, we remark that $L_1 (D_6,w_6)$ and $L_1 (D_6,w_1)$ are fermionic modules in \cite{Lee:2024fxa}. This is consistent with fact that  $L_1 (D_6,w_6)\otimes L(-3/5,3/4)$ in \eqref{Wiso1} is odd for $\sigma$ of the $\ZZ_2=\{1,\sigma\}$ automorphism group of $\cW_{-3}(E_7,f_\theta)$.

The $\cW_{-2}(E_6,f_\theta)$ algebra is a bit more complicated as $E_6$ has a nontrivial outer automorphism group $\ZZ_2$. It has six Ramond twisted irreducible modules as follows
\begin{align}\nonumber
     R_0& =L_1 (A_5)\otimes L(-3/5,-1/20)\oplus L_1 (A_5,w_3)\otimes L(-3/5,1/5),\\ \nonumber
    R_1& =L_1 (A_5)\otimes L(-3/5,3/4)\oplus L_1 (A_5,w_3)\otimes L(-3/5,0),\\\nonumber
    R_3& =L_1 (A_5,w_1)\otimes L(-3/5,0)\oplus L_1 (A_5,w_4)\otimes L(-3/5,3/4),\\\nonumber
    R_4& =L_1 (A_5,w_1)\otimes L(-3/5,1/5)\oplus L_1 (A_5,w_4)\otimes L(-3/5,-1/20),\\[-0.5mm] \nonumber
    \overline{R}_3& =L_1 (A_5,w_5)\otimes L(-3/5,0)\oplus L_1 (A_5,w_2)\otimes L(-3/5,3/4),\\[-0.5mm]
    \overline{R}_4& =L_1 (A_5,w_5)\otimes L(-3/5,1/5)\oplus L_1 (A_5,w_2)\otimes L(-3/5,-1/20).\nonumber
\end{align}
Note there exist two pairs of complex conjugated modules. The characters of $R_0,R_1,R_2,R_3$ coincide with the characters of $L_1(\Ah)$ with conformal dimension $0,4/5,7/15,2/3$ respectively \cite{Lee:2024fxa}. As vector valued modular form, they can be realized by $\T_{7}$ Hecke image of a simple current extension of $\mathrm{Vir}_{6,5}$ that is the 3-state Potts model \cite{Lee:2024fxa}. The complex conjugation matches with the degeneracy of characters in \cite{Lee:2024fxa}.

Again we use a simple current of $L_1(A_5)$ to recover the irreducible modules of $\cW_{-2}(E_6,f_\theta)$. The $L_1(A_5)$ has a $\ZZ_2$-simple current $L_1(A_5,w_3)$ which induces a $\ZZ_2$ automorphism for the six irreducible modules of $L_1(A_5)$. To be precise, it induces the following exchanges
$$
L_1(A_5)\leftrightarrow L_1(A_5,w_3),\quad L_1(A_5,w_2)\leftrightarrow L_1(A_5,w_5),\quad L_1(A_5,w_1)\leftrightarrow L_1(A_5,w_4).
$$
The six irreducible modules of $\cW_{-2}(E_6,f_\theta)$ are recovered by the above exchanges. For example, $R_1$ is mapped to $N_0\cong \cW_{-2}(E_6,f_\theta)$ in \eqref{Wiso1}. Note the $L_1 (A_5,w_1),L_1 (A_5,w_3)$ and $L_1 (A_5,w_5)$ are fermionic modules in \cite{Lee:2024fxa}. This is consistent with fact that  $L_1 (A_5,w_3)\otimes L(-3/5,3/4)$ in \eqref{Wiso1} is odd for $\sigma$ of the $\ZZ_2=\{1,\sigma\}$ automorphism group of $\cW_{-2}(E_6,f_\theta)$. 

Remarkably, the above correspondence between $\cW_{-h^\vee/6}(\mg,f_\theta)$ and $L_1(\mg_I)$ for $\mg=E_{6,7,8}$ can be generalized to $\cW_{-h^\vee/6+1}(\mg,f_\theta)$ and $L_2(\mg_I)$. 
We find that (i) the Ramond-twisted irreducible modules  of $\cW_{-h^\vee/6+1}(\mg,f_\theta)$ coincide with the irreducible modules  of $L_2(\mg_I)$, (ii) the irreducible modules  of $\cW_{-h^\vee/6+1}(\mg,f_\theta)$ are related to the Ramond-twisted ones by a $\ZZ_2$ automorphism of $L_2(\mg_*)$ induced by the simple current $L_2 (\mg_*,2w_*)$. This simple current of $L_2(\mg_*)$ always has  conformal dimension $3/2$. 
The coset constructions for $\cW_{-h^\vee/6+1}(\mg,f_\theta)$ and $L_2(\mg_I)$ are more complicated than \eqref{coset1}, and have been presented in \eqref{cosetWE6}, \eqref{cosetWE7} and \eqref{cosetWE8} in the introduction.

Our aim is to find all irreducible modules  of $\cW_{-h^\vee/6+1}(\mg,f_\theta)$. To achieve that, we first need to find all irreducible modules of $L_2(\mg_I)$. Fortunately, there have been some recent progress on the characters of $L_2(\Eh)$ \cite{Duan:2022ltz,Lee:2023owa} and $L_2(\Dh)$ \cite{Lee:2024fxa} with the help of Hecke operator \cite{Harvey:2018rdc} and fermionic Hecke operator \cite{Lee:2022yic}. We follow this approach to find the characters of $L_2(\Ah)$ by Hecke operator. The Hecke operator $\T_p$ proposed by Harvey--Wu in \cite{Harvey:2018rdc} is naturally defined for vector-valued modular forms of weight $0$ on $SL(2,\ZZ)$ and often maps the characters of a 2d RCFT with (effective) central charge $c$ to the characters of another 2d RCFT with central charge $pc$. It also maps the (effective) conformal dimension $h_i$ to $p h_i\!\!\mod \ZZ$. To confirm a vector-valued modular form describes the characters of an intermediate vertex subalgebra, many checks can be performed with the knowledge on intermediate Lie algebras such the module dimensions and quadratic Casimir invariants that give the conformal dimensions. Relevant information of intermediate Lie algebras for the current work has been obtained in \cite{Lee:2023owa,Lee:2024fxa} and cross-checked with Vogel's universal Lie algebra \cite{vogel1999universal} and various exceptional series \cite{LMseries}. 
For example, the intermediate Lie algebras $\Eh$ and $\Dh$ lie on Vogel's projective plane. The $\Eh,\Dh$ and $\Ah$ lie in Cvitanovi\'c--Deligne, sub and Severi exceptional series respectively.

By Hecke operator we can compute the characters of $L_2(\mg_I)$ to arbitrary high $q$ degrees, along with the $S$-matrices.  
With these  at hand, we establish the coset construction $L_2(\mg_I)/L_2(\mg_*)$ and find the decompositions to $L_2(\mg_*)$ modules. By the correspondence between $L_2(\mg_I)$ and $\cW_{-h^\vee/6+1}(\mg,f_\theta)$, we have all Ramond-twisted irreducible modules of $\cW_{-h^\vee/6+1}(\mg,f_\theta)$. In the end, by the $\ZZ_2$ automorphism induced by the simple current $L_2 (\mg_*,2w_*)$, we recover all irreducible modules of $\cW_{-h^\vee/6+1}(\mg,f_\theta)$, in particular the adjoint ones presented in \eqref{isoE6}, \eqref{isoE7} and \eqref{isoE8}. In the following sections, we will discuss this procedure one by one for $\mg=E_{6,7,8}$.

Most recently, a denominator identity for the minimal $\cW$-algebras for $\mg=D_4,E_{6,7,8}$ is given in \cite[Section 14]{Kac:2024kvv}. It would be interesting to connect this denominator formula and character computation in this work. We remark that such denominator formula may be regarded as the Macdonald--Weyl denominator identity for intermediate Lie algebras. 
It is also interesting to consider $\cW_k(\mathfrak{g},f_\theta)$ for $\mg=F_4$ and $k=-1/2$ and see if the coset constructions of type A and B still hold for $\cW_{-1/2}(F_4,f_\theta)$. Here the parameter $m=24/(h^\vee-6)=8$. The $\cW_{-3/2}(F_4,f_\theta)$ has been studied in \cite{Kawasetsu:2018irs,Kawasetsu:2018tzs}, which has 4 irreducible modules. We expect $\cW_{-1/2}(F_4,f_\theta)$ has 10 irreducible modules and can be decomposed to $L_2(C_3)\otimes M_4$ modules, where $M_4$ is an extension of $\cB\mathrm{Vir}^{N=1}_{34,12}$.

\section{$\cW_{-4}(E_8,f_\theta)$}\label{sec:E8}
The minimal $\cW$-algebra $\cW_{-4}(E_8,f_\theta)$ has central charge $c={154}/{13}$. The affine VOA $L_2(E_7)$ has central charge $c=133/10$. The intermediate vertex subalgebra $L_2(\Eh)\subset L_2(E_8)$ has central charge $c=190/13$. We find the following coset constructions
\begin{align}\label{WE8cons}
    \frac{\cW_{-4}(E_8,f_\theta)}{ L_2(E_7)}=    \cB \mathrm{Vir}_{13,5}^{N=1},
\end{align}
and
\begin{align}\label{Ehcons}
    \frac{L_2(\Eh)}{ L_2(E_7)}=  \left(\cB \mathrm{Vir}_{13,5}^{N=1}\right)_{\rm eff}.
\end{align}
In the following, we verify these cosets and use them to determine all irreducible modules and Ramond-twisted ones of $\cW_{-4}(E_8,f_\theta)$.

\subsection{Coset A}

We first recall some basic properties of $N=1$ supersymmetric minimal models $\mathrm{Vir}^{N=1}_{p,q}$ \cite{Friedan:1983xq,Friedan:1984rv,Bershadsky:1985dq}, which are the simplest rational and $C_2$-cofinite vertex operator superalgebras with only super Virasoro symmetry. They are labeled by two positive integers $(p,q)$ satisfying $2\le q\le p-2$, $p-q\in 2\ZZ$ and $\gcd(\frac{p-q}{2},p)=1$. The central charge of $\mathrm{Vir}^{N=1}_{p,q}$ is 
\begin{align}
    c=\frac{3}{2}\Big(1-\frac{2(p-q)^2}{pq}\Big).
\end{align}
The $\mathrm{Vir}^{N=1}_{p,q}$ has two types of irreducible modules called Neveu-Schwarz (NS) and Ramond (R) modules. Their conformal dimensions are given by
\begin{align}
h_{r,s}=\frac{(pr-qs)^2-(p-q)^2}{8pq}+\frac{2\epsilon_{r-s}-1}{16},\qquad \epsilon_a =
\begin{cases}
  \frac12\quad  a\in 2\ZZ ,\\
  1 \quad a\in 2\ZZ+1.
  \end{cases}
\end{align}
Here $r=1,2,\ldots,q-1$, $s=1,2,\ldots,p-1$ and $qs\le pr$. The $h_{r,s}$ with $r-s\in 2\ZZ$ are the NS conformal dimensions, while those with $r-s\in 2\ZZ+1$ are the R conformal dimensions. The SVOA 
$\mathrm{Vir}^{N=1}_{p,q}$ is unitary if and only if $p=q+2$. For non-unitary $\mathrm{Vir}^{N=1}_{p,q}$, it is often useful to consider the effective description to make the effective central charge and effective conformal dimensions non-negative. Denote the lowest conformal dimension as $h_{\rm min}$, then the effective central charge is defined as $c_{\rm eff}=c-24 h_{\rm min}$, and effective conformal dimensions are shifted accordingly by $h_{r,s}^{\rm eff}=h_{r,s}- h_{\rm min}$. It is easy to show for $\mathrm{Vir}^{N=1}_{p,q}$, 
\begin{align}\label{ceff}
    c_{\rm eff}=\frac{3}{2}\Big(1-\frac{2(p-q)^2}{pq}\Big),
\end{align}
and 
\begin{align}
h^{\rm eff}_{r,s}=\frac{(pr-qs)^2-4}{8pq}+\frac{2\epsilon_{r-s}-1}{16},\qquad \epsilon_a =
\begin{cases}
  \frac12\quad  a\in 2\ZZ ,\\
  1 \quad a\in 2\ZZ+1.
  \end{cases}
\end{align}
The character formulas of the irreducible Neveu-Schwarz and Ramond  modules are given in \cite{Goddard:1986ee}.  

The coset constructions and Hecke operators in the current work involve the bosonizations of $\mathrm{Vir}^{N=1}_{p,q}$ and their further extensions. 
Some simplest examples of the bosonization of supersymmetric minimal models include 
\begin{align}
    \cB\mathrm{Vir}^{N=1}_{5,3}\cong \mathrm{Vir}_{5,4}
\end{align}
for the unitary case and
\begin{align}
    \cB\mathrm{Vir}^{N=1}_{8,2}\cong \mathrm{Vir}_{8,3}
\end{align}
for the non-unitary case. 
In general, $\cB\mathrm{Vir}^{N=1}_{p,q}$ are some rational VOAs which are not necessarily isomorphic familiar VOAs. They usually have a large number of irreducible modules. To the best of our knowledge, the bosonizations of $\mathrm{Vir}^{N=1}_{p,q}$ used in the current paper with $(p,q)=(13,5),(16,6),(22,8),(132,2)$ are not studied before. Suppose one of these $\mathrm{Vir}^{N=1}_{p,q}$ has $K$ irreducible NS modules, then $\cB\mathrm{Vir}^{N=1}_{p,q}$ has $3K$ irreducible modules if $K$ is even, or $3K+1$ irreducible modules if $K$ is odd. As a Neveu-Schwarz module is $\ZZ/2$-graded, the integer and half-integer parts become two irreducible modules of $\cB\mathrm{Vir}^{N=1}_{p,q}$, which we often denote as $\phi$ and $\varphi$. On the other hand, Ramond sector is $\ZZ$-graded, a Ramond module becomes directly an irreducible module of $\cB\mathrm{Vir}^{N=1}_{p,q}$ except when the exponent $-c/24+h_{\rm R}=0$. In this case, a constant $\rm\widetilde{R}$ sector comes to play and the Ramond  module with $\alpha=0$ splits to two irreducible modules of $\cB\mathrm{Vir}^{N=1}_{p,q}$ with characters $(\chi_{\rm R}\pm 1)/2$. This happens when $(p,q)=(16,6),(22,8),(132,2)$ in the current paper. We denote the $K$ or $K+1$ irreducible modules obtained from the Ramond  sector by $\psi$.

We now discuss $\mathrm{Vir}_{13,5}^{N=1}$ and its bosonization. The $N=1$ minimal model $\mathrm{Vir}^{N=1}_{13,5}$ has central charge $-{189}/{130}$ and effective central charge ${171}/{130}$. 
It has 12 Neveu-Schwarz modules with conformal dimensions from small to big as
\begin{align}\label{SM135NS}
 \left\{-\frac{3}{26},-\frac{6}{65},-\frac{7}{130},0,\frac{2}{13},\frac{33}{130},\frac{24}{65},\frac{21}{26},\frac{64}{65},\frac{24}{13},\frac{21}{10},\frac{85}{26}\right\}   .
\end{align}
The 12 Ramond modules with conformal dimensions from small to big are 
\begin{align}\label{SM135R}
 \left\{-\frac{61}{1040},-\frac{9}{208},\frac{7}{208},\frac{99}{1040},\frac{179}{1040},\frac{103}{208},\frac{659}{1040},\frac{63}{80},\frac{279}{208},\frac{1619}{1040},\frac{535}{208},\frac{67}{16}\right\}   .
\end{align}
There is no R conformal dimension with zero exponent, thus  $\cB\mathrm{Vir}_{13,5}^{N=1}$ has in total $12\times 3=36$ irreducible modules. The $12$ irreducible modules of $\cB\mathrm{Vir}_{13,5}^{N=1}$ in $\phi$ sector has conformal dimension as in \eqref{SM135NS}. The $12$ irreducible modules in $\varphi$ sector have conformal dimensions as those in  \eqref{SM135NS} added by $1/2$, except the $0$ is added by $3/2$.  The $12$ irreducible modules in $\psi$ sector have conformal dimensions as in \eqref{SM135R}.  
The effective conformal dimensions of $\cB\mathrm{Vir}_{13,5}^{N=1}$ are the conformal dimensions added by $3/22$.

The characters of $L_2(\Eh)$ were found in \cite{Duan:2022ltz} by Hecke operator $\T_{19}$ on $\mathrm{Vir}_{13,2}$, where the effective central charge $10/13$ of $\mathrm{Vir}_{13,2}$ are mapped to $190/13$ as required. The six characters of $L_2(\Eh)$ are presented in \cite[Equation (7.2)]{Duan:2022ltz}. We use the characters to verify the coset construction \eqref{Ehcons}. For example, for the vacuum character $\chi_0$ of $L_2(\Eh)$, we find
the character relation of coset \eqref{Ehcons} as 
\begin{align}
   \chi_0 & = \chi^{\cB}_0 \chi_0^{E}+\chi^{\cB}_{\frac{23}{80}} \chi_{\frac{57}{80}}^{E}+\chi^{\cB}_{\frac{11}{10}} \chi_{\frac{9}{10}}^{E}+\chi^{\cB}_{\frac{43}{16}} \chi_{\frac{21}{16}}^{E}+\chi^{\cB}_{\frac{8}{5}} \chi_{\frac{7}{5}}^{E}+\chi^{\cB}_{\frac{1}{2}} \chi_{\frac{3}{2}}^{E}. 
\end{align}
Here ${\cB}$ is short for the effective $\cB\mathrm{Vir}_{13,5}^{N=1}$, while ${E}$ is short for $L_2(E_7)$. The weight-1 space is contributed from the first two terms in the right hand side, which gives the long-known module decomposition $\bf 190=\bf 133+56+1$ for $E_7\subset \Eh$. Similarly we find and verify the character relations of coset \eqref{Ehcons} for all other five characters of $L_2(\Eh)$. 

Owing to the correspondence between $\cW_{-4}(E_8,f_\theta)$ and $L_2(\Eh)$, the above character relations can be translated to the character relations for the Ramond-twisted irreducible modules of $\cW_{-4}(E_8,f_\theta)$ in coset \eqref{WE8cons}. We only need to replace ${\cB}$ from the effective $\cB\mathrm{Vir}_{13,5}^{N=1}$ to the original $\cB\mathrm{Vir}_{13,5}^{N=1}$ (denoted as $M_8$) and change the effective conformal dimension to the original ones. In this way, we determine the decomposition to $L_2(E_7)\otimes M_8$ modules for all six Ramond-twisted irreducible modules. The results are summarized in Table \ref{tab:WE8Ramondirr}, where the character of $R_i$ is $\chi_{\mu_i}$ in \cite[Equation (3.3)]{Lee:2023owa}. The $\omega_i$ are the analogy of highest-weights for $\Eh$ irreducible modules such that $\chi_{\mu_i}$ can be regarded as the character of $L_2(\Eh,\omega_i)$ \cite{Lee:2023owa}.

\begin{table}[ht]
\caption{All Ramond-twisted irreducible modules of $\cW_{-4}(E_8,f_\theta)$. The conformal dimensions $h_{ij}$ are shown for each module $R_i=\bigoplus_j L_2(E_7,\lambda_j)\otimes M_8(h_{ij})$. }
\def\arraystretch{1.1}
    \centering
    \begin{tabular}{c|c|c|c|c|c|c|c|ccc} \hline
$\omega_i$ & $R_i$   &  $0$  &  $ 2w_6$    &   $ w_1 $   & $ w_5$   &   $w_6 $    &  $ w_7$   &  $ \mu_i$
 \\ \hline    
 $0$  &$R_0$ &     $ -3/26   $   &   $  5/13  $   &    $ 64/65   $   &     $  193/130  $  &    $  179/1040  $   &    $  535/208  $  &    $   0 $          \\ 
$ 2w_6$    &$R_1$ &      $  3/2  $   &   $ 0   $   &    $   13/5 $   &     $  21/10  $  &    $  63/80  $   &    $  67/16  $     &    $  21/13  $      \\ 
$ w_1 $   &$R_2$ &   $  21/26  $   &   $ 17/13   $   &    $  -6/65  $   &     $  53/130  $  &    $ 99/1040   $   &    $  103/208  $    &    $ 12/13   $     \\ 
$ w_5$   &  $R_3$ &     $  61/26  $   &   $  24/13  $   &    $  29/65  $   &     $  -7/130  $  &    $  659/1040  $   &    $  7/208  $   &    $  19/13  $       \\ 
 $w_6 $    & $R_{4}$ &     $   17/26 $   &   $  2/13  $   &    $  49/65  $   &     $  33/130  $  &    $  -61/1040  $   &    $   279/208 $   &    $ 10/13   $      \\ 
$ w_7$   & $R_{5}$ &      $  85/26  $   &   $  49/13  $   &    $ 24/65   $   &     $  113/130  $  &    $  1619/1040  $   &    $ -9/208   $    &    $ 18/13   $       \\ 
 \hline
     \end{tabular}
    \label{tab:WE8Ramondirr}
\end{table}

\subsection{Irreducible modules of $\cW_{-4}(E_8,f_\theta)$}

We now recover all irreducible modules of $\cW_{-4}(E_8,f_\theta)$ from the Ramond-twisted ones given in Table \ref{tab:WE8Ramondirr}. 
The non-identity simple current of $L_2(E_7)$ is well-known as $L_2(E_7,2w_6,3/2)$. 
The simple current induces a $\ZZ_2$ automorphism for the irreducible modules of $L_2(E_7)$. To be precise, it induces the following exchanges
\begin{align}\nonumber
L_2(E_7,0,0)\leftrightarrow L_2(E_7,2w_6,3/2) ,\qquad \ L_2(E_7,w_1,9/10 )\leftrightarrow L_2(E_7,w_5 , 7/5) .
\end{align}
The remaining two irreducible modules $L_2(E_7,w_6,57/80)$ and $L_2(E_7,w_7,21/16) $ fusion to themselves.

The irreducible modules of $\cW_{-4}(E_8,f_\theta)$ are related to the Ramond-twisted ones by the above $\ZZ_2$ automorphism for $L_2(E_7)$ modules. We then easily obtain all six irreducible modules $N_0,\dots,N_5$ and their decompositions to $L_2(E_7)\otimes M_8$ irreducible modules and summarize the results in Table \ref{tab:WE8irr}. For example, $N_0\cong \cW_{-4}(E_8,f_\theta)$ has been shown in \eqref{isoE8}. The $\cW_{-4}(E_8,f_\theta)$ is $\ZZ_2$-rational with automorphism group $\ZZ_2=\{1,\sigma\}$. 
The first four and last two columns in Table \ref{tab:WE8irr} contribute to the even and odd parts of $\cW_{-4}(E_8,f_\theta)$ respectively. This means $\sigma(b)=b$ and $\sigma(f)=-f$ for any $b $ in $L_2(E_7)\otimes M_8$ modules in the first four columns and $f$ in $L_2(E_7)\otimes M_8$ modules in the last two columns. Note the last two weights $w_6$ and $w_7$ are referred to as fermionic weights in \cite{Lee:2023owa}. 

\begin{table}[ht]
\caption{All irreducible modules of $\cW_{-4}(E_8,f_\theta)$. The conformal dimensions $h_{ij}$ are shown for each irreducible module $N_i=\bigoplus_j L_2(E_7,\lambda_j)\otimes M_8(h_{ij})$.}
\def\arraystretch{1.1}
    \centering
    \begin{tabular}{c|c|c|c|c|c|ccccc} \hline
 $\lambda$   &  $0$  &  $ 2w_6$    &   $ w_1 $   & $ w_5$   &   $ w_6$    &  $w_7 $   
 \\ \hline    
$N_0$ &     $  0  $   &   $  3/2  $   &    $  21/10  $   &     $  13/5  $  &    $  63/80  $   &    $  67/16  $          \\ 
$N_1$ &      $  49/13  $   &   $  85/26  $   &    $  113/130  $   &     $  24/65  $  &    $  1619/1040  $   &    $ -9/208   $         \\ 
$N_2$ &   $  24/13  $   &   $  61/26  $   &    $  -7/130  $   &     $ 29/65   $  &    $  659/1040  $   &    $  7/208  $       \\ 
$N_3$ &     $  2/13  $   &   $  17/26  $   &    $  33/130  $   &     $49/65    $  &    $  -61/1040  $   &    $  279/208  $        \\ 
$N_4$ &     $  17/13  $   &   $  21/26  $   &    $  53/130  $   &     $ -6/65   $  &    $  99/1040  $   &    $  103/208  $       \\ 
$N_5$ &      $  5/13  $   &   $ -3/26   $   &    $   193/130 $   &     $  64/65  $  &    $  179/1040  $   &    $ 535/208   $         \\ 
 \hline
     \end{tabular}
    \label{tab:WE8irr}
\end{table}
 
\subsection{Coset B}
The coset construction
$$
\frac{L_2(E_8)}{L_2(\Eh)}=\mathrm{Vir}^{\rm eff}_{13,4}
$$
was given in \cite[Section 3]{Lee:2023owa}. The character relations for the cosets are presented in \cite[Equation 3.9]{Lee:2023owa}. As in this case, the characters and irreducible modules are one-to-one corresponding to each other, we omit the module decompositions here. This coset shows that $L_2(\Eh)$ is an intermediate vertex subalgebra of $L_2(E_8)$.

\section{$\cW_{-2}(E_7,f_\theta)$}\label{sec:E7}
The $\cW_{-2}(E_7,f_\theta)$ algebra has central charge $c= {75}/{8}.$ The affine VOA $L_2(D_6)$ has central charge $c=11.$ The intermediate vertex subalgebra $L_2(\Dh)\subset L_2(E_7)$ has central charge $c={99}/{8}.$ We find the following coset construction
\begin{align}\label{WE7cons}
    \frac{\cW_{-2}(E_7,f_\theta)}{ L_2(D_6)}=    \cB \mathrm{Vir}_{16,6}^{N=1},
\end{align}
and
\begin{align}\label{cosDh}
    \frac{L_2(\Dh)}{ L_2(D_6)}=  \left(\cB \mathrm{Vir}_{16,6}^{N=1}\right)_{\rm eff}.
\end{align}
We will  verify  and make use of these cosets to determine all irreducible modules and Ramond-twisted ones of $\cW_{-2}(E_7,f_\theta)$.

\subsection{Coset A}

Following the same approach to $\cW_{-4}(E_8,f_\theta)$ in the last section, to find the Ramond-twisted modules of $\cW_{-2}(E_7,f_\theta)$, we first discuss the characters and module decompositions of $L_2(\Dh)$. Many useful results on $L_2(\Dh)$ have been obtained recently in \cite{Lee:2024fxa}. 
Notably, both $L_2(D_6)$ and $L_2(\Dh)$ allow fermionization. Here the fermionization $\cF$ of an affine VOA $L_k(\mg)$ means that there exists a $\ZZ_2$-simple current of conformal dimension $3/2$ which induces a simple current extension of $L_k(\mg)$ to a $N=1$ SVOA. The $\cF L_2(D_6)$ has 4 Neveu-Schwarz irreducible modules with conformal dimensions $0, {11}/{16}, {5}/{6},1$,  
% \begin{align}
%     0,\frac{11}{16},\frac{5}{6},1.
% \end{align} 
while the  
$\cF L_2(\Dh)$ has 4 Neveu-Schwarz irreducible modules with conformal dimensions $0, 3/4, 7/8, 9/8$ \cite{Lee:2024fxa}.

The affine VOA $L_2(D_6)$ has 13 irreducible modules with conformal dimensions
\begin{align}
 0,\frac{11}{24}, \left(\frac{11}{16}\right)_2, \frac56,1,\frac98,\left(\frac{19}{16}\right)_2,\frac43,\frac{35}{24},\left(\frac32\right)_2 .
\end{align}
Here $()_2$ means degeneracy 2 of the characters, as $D_6$ has an outer automorphism group $\ZZ_2$. 
The 8 irreducible modules that form NS modules of $\cF L_2(D_6)$ have the following four NS pairs of conformal dimensions
\begin{align}
 \left(0,\frac32\right), \left(\frac{11}{16},\frac{19}{16}\right) ,\left(\frac56,\frac43\right),\left(1,\frac32\right).
\end{align}
The remaining 5 irreducible modules with conformal dimensions $ {11}/{24},  {11}/{16},  9/8, {19}/{16} $ and $ {35}/{24}$ form the four Ramond-twisted modules of $\cF L_2(D_6)$. 

On the other hand, the intermediate vertex subalgebra $L_2(\Dh)$ has 13 irreducible modules with conformal dimensions \cite{Lee:2024fxa}
\begin{align} 
    \left\{0,\frac{33}{64},\frac{45}{64},\frac{3}{4},\frac{7}{8},\frac{9}{8},\frac{77}{64},\frac{5}{4},\frac{81}{64},\frac{11}{8},\frac{3}{2},\frac{97}{64},\frac{13}{8}\right\}.
\end{align}
The 8 irreducible modules that form NS modules of $\cF L_2(\Dh)$ have the following four NS pairs of conformal dimensions
\begin{align}
 \left(0,\frac32\right), \left(\frac{3}{4},\frac{5}{4}\right) ,\left(\frac78,\frac{11}{8}\right),\left(\frac98,\frac{13}{8}\right).
\end{align}
The rest 5 irreducible modules with conformal dimensions $ {33}/{64},{45}/{64},  {77}/{64},{81}/{64}$ and $ {97}/{64}$ form the four Ramond-twisted modules of $\cF L_2(\Dh)$. 
The supercharacters of $\cF L_2(\Dh)$ were given in \cite[Section 4.3.2.]{Lee:2024fxa}. As modular forms of congruence subgroup of level $2$, they have appeared in the study on fermionic modular bootstrap  \cite{Duan:2022kxr} and fermionic Hecke operator \cite{Lee:2022yic}. With the NS and R characters, it is easy to recover the 13 characters of $L_2(\Dh)$. 
Given the importance of these characters, we present them in $q$-series  explicitly as follows:
\begin{align}\nonumber
    \chi_0&=q^{-\frac{33}{64}}(1+99 q+5049 q^2+86889 q^3+932085 q^4+7446186 q^5+\dots),\\[+1mm]\nonumber
     \chi_{\frac{33}{64}}&= 12+1188 q+30096 q^2+393184 q^3+3567960 q^4+25460424 q^5+ \dots,\\ \nonumber
     \chi_{\frac{45}{64}}&=q^{\frac{3}{16}}( 44+4356 q+93456 q^2+1123848 q^3+9681804 q^4+\dots  ),\\ \nonumber
    \chi_{\frac{3}{4}}&=q^{\frac{15}{64}}( 33+3267 q+67518 q^2+796125 q^3+6774273 q^4+\dots ),\\ \nonumber
    \chi_{\frac{7}{8}}&=q^{\frac{23}{64}}( 99+5874 q+111078 q^2+1245618 q^3+10266828 q^4+\dots ),\\ \nonumber
    \chi_{\frac{9}{8}}&=q^{\frac{39}{64}}( 77+2574 q+42075 q^2+429771 q^3+3339072 q^4+\dots ),\\ \label{chiD0}
     \chi_{\frac{77}{64}}&=q^{\frac{11}{16}}(  252+9240 q+141372 q^2+1412532 q^3+10770188 q^4+\dots  ),\\ \nonumber
    \chi_{\frac{5}{4}}&=q^{\frac{47}{64}}(  495+16280 q+244035 q^2+2399067 q^3+18107529 q^4+\dots ),\\   \nonumber
    \chi_{\frac{81}{64}}&=q^{\frac{3}{4}}( 616+20592 q+304776 q^2+2983728 q^3+22438152 q^4+\dots,\\\nonumber
    \chi_{\frac{11}{8}}&=q^{\frac{55}{64}}( 945+27918 q+389895 q^2+3689928 q^3+27102438 q^4+\dots),\\    \nonumber
     \chi_{\frac32}&=q^{\frac{63}{64}}(891+22572 q+297891 q^2+2713546 q^3+19428894 q^4+\dots),\\\nonumber
    \chi_{\frac{97}{64}}&=1188 q+30096 q^2+393184 q^3+3567960 q^4+25460424 q^5+\dots,\\ \nonumber
    \chi_{\frac{13}{8}}&=q^{\frac{71}{64}}( 495+11253 q+140184 q^2+1232748 q^3+8610151 q^4+\dots ).
\end{align}
All conformal dimensions in \eqref{chiD0} are consistent with the quadratic Casimir invariants of $\Dh$ irreducibles in \cite[Table 6]{Lee:2024fxa}, and all leading Fourier coefficients are the irreducible module dimensions of $\Dh$ in \cite[Table 6]{Lee:2024fxa}. Note that $\chi_{\frac{97}{64}}-\chi_{\frac{33}{64}}=12 $ as expected from $N=1$ supersymmetry. 

Now we discuss the coset construction \eqref{cosDh}.
The $\mathrm{Vir}_{16,6}^{N=1}$ has central charge $-{13}/{8}$  and   effective central charge ${11}/{8}$. It has 19 Neveu-Schwarz irreducible modules with conformal dimensions
\begin{align}\label{phi166}
    \left\{-\frac{1}{8},-\frac{7}{64},-\frac{1}{12},-\frac{3}{64},0,\frac{1}{8},\frac{13}{64},\frac{7}{24},\frac{25}{64},\frac{3}{4},\frac{57}{64},\frac{25}{24},\frac{7}{4},\frac{125}{64},\frac{13}{6},\frac{25}{8},\frac{217}{64},\frac{39}{8},7\right\}  ,
\end{align}
and 19 Ramond modules  with conformal dimensions
\begin{align}\label{psi166}
    \left\{-\frac{13}{192},-\frac{1}{16},-\frac{3}{64},\frac{1}{64},\frac{1}{16},\frac{23}{192},\frac{3}{16},\frac{29}{64},\frac{9}{16},\frac{131}{192},\frac{13}{16},\frac{81}{64},\frac{23}{16},\frac{311}{192},\frac{157}{64},\frac{43}{16},\frac{257}{64},\frac{69}{16},\frac{381}{64}\right\} .
\end{align}
The bosonization $\cB\mathrm{Vir}^{N=1}_{16,6}$ has in total $3\times 19+1=58$ irreducible modules. In this case, the R character with conformal dimension $- {13}/{192}$ has exponent zero and splits to two bosonic characters. The 58 conformal dimensions of $\cB\mathrm{Vir}^{N=1}_{16,6}$ are (i) $\phi$ sector: all dimensions in \eqref{phi166}, (ii) $\varphi$ sector: all dimensions in \eqref{phi166} added by $1/2$ except the $0$ added by $3/2$, (iii) $\psi$ sector: all dimensions in \eqref{psi166} with one more $179/192$. The effective conformal dimensions are given by all conformal dimensions added by $1/8$. 

Now we verify the coset \eqref{cosDh} by checking the character relations to sufficiently high $q$ degrees. For example, for the vacuum character of $L_2(\Dh)$, i.e., the $\chi_0$ in \eqref{chiD0}, we find 
\begin{align}\label{Dhchi0}
 \chi_0 & = \chi^{\cB}_0 \chi_0^{D}+{\chi^{\cB}_{\frac{ 5}{16 }} \chi_{\frac{ 11  }{ 16  }}^{D}}+\chi^{\cB}_{ \frac{7}{6}} \chi_{\frac{5}{6}}^{D}+\chi^{\cB}_{5} \chi_{1}^{D}+{\chi^{\cB}_{\frac{ 45}{ 16}} \chi_{\frac{  19 }{ 16  }}^{D}}+\chi^{\cB}_{\frac{ 5}{3 }} \chi_{\frac{  4 }{ 3  }}^{D}+\chi^{\cB}_{\frac{1 }{2 }} \chi_{\frac{   3}{  2 }}^{D}+\chi^{\cB}_{\frac{11 }{2 }} \chi_{\frac{   3}{2   }}^{D}   . 
    \end{align}
Here $ D$ is short for $ {L_2(D_6)}$ and $ {\cB}$ is short for  $(\cB\mathrm{Vir}^{N=1}_{16,6})_{\rm eff}$. The weight-1 space of $\chi_0$ is contributed from the first two terms in the right hand side of \eqref{Dhchi0}, which gives precisely the module decomposition $\bf 99=\bf 66+\overline{32}+1$ for $D_6\subset \Dh$ in \cite{Lee:2024fxa}. For the NS partner of the vacuum which has conformal dimension $3/2$, i.e., the supersymmetry generator, we find 
\begin{align}\label{Dhchi32}
   \chi_{\frac32}  = \chi^{\cB}_{ \frac{11}{2}} \chi_0^{D}+\chi^{\cB}_{\frac{ 45 }{  16}} \chi_{\frac{ 11  }{ 16  }}^{D}+\chi^{\cB}_{ \frac{ 5}{3 }} \chi_{\frac{5}{6}}^{D}+\chi^{\cB}_{\frac12  } \chi_{1}^{D}+\chi^{\cB}_{\frac{  5}{16  }} \chi_{\frac{  19 }{ 16  }}^{D}+\chi^{\cB}_{\frac{ 7 }{ 6 }} \chi_{\frac{  4 }{ 3  }}^{D}+\chi^{\cB}_{0} \chi_{\frac{   3}{  2 }}^{D}+\chi^{\cB}_{5} \chi_{\frac{   3}{2   }}^{D}  ,
 \end{align}
The weight-0 space is contributed from the forth, fifth and seventh terms in the right hand side of \eqref{Dhchi32}, which gives precisely the module decomposition $\bf 891=\bf 462 + \overline{352} + 77$ for $D_6\subset \Dh$ in \cite{Lee:2024fxa}.
Compared with \eqref{Dhchi0}, it is easy to observe that the $\cB$ coefficients do not change, while $L_2({D}_6)$ module changes to its NS partner. This persists for all four NS pairs of characters of $L_2(\Dh)$.

By a  large amount of computation, we find the character relations of coset \eqref{cosDh} for all 13 characters of $L_2(\Dh)$ in \eqref{chiD0}. Then by carefully comparing with the module decompositions under $D_6\subset \Dh$ in \cite{Lee:2024fxa}, we can further upgrade the character relations to module decompositions for $L_2(\Dh)$. As the characters of $L_2(\Dh)$ coincide with the characters of Ramond-twisted irreducible modules of $\cW_{-2}(E_7,f_\theta)$, we can translate the module decompositions  to those of $\cW_{-2}(E_7,f_\theta)$ in coset \eqref{WE7cons} by simply replacing $(\cB\mathrm{Vir}^{N=1}_{16,6})_{\rm eff}$ to $\cB\mathrm{Vir}^{N=1}_{16,6}$ (denoted as $M_7$).  In the end, we obtain all 13 Ramond-twisted irreducible modules of $\cW_{-2}(E_7,f_\theta)$ and their decomposition to $L_2(D_6)\otimes M_7$ modules in  coset construction \eqref{WE7cons}. The results are  summarized in the Table \ref{tab:WE7Ramondirrbosonic} and \ref{tab:WE7Ramondirrfermionic} for irreducibles that form the NS and R sector of $\cF L_2(\Dh)$ respectively. The character of $R_i$ is $\chi_{\mu_i}$ in \eqref{chiD0}. The $(R_0,R_3)$, $(R_1,R_2)$, $(R_4,R_5)$, $(R_6,R_7)$ in Table  \ref{tab:WE7Ramondirrbosonic} form NS pairs. Each Ramond-twisted irreducible module $R_i$ can be regarded as an irreducible module $L_2(\Dh,\omega_i,\mu_i)$  in the sense of \cite{Lee:2024fxa}.

\begin{table}[ht]
\caption{All Ramond-twisted irreducible modules of $\cW_{-2}(E_7,f_\theta)$ that form NS pairs. The $h_{ij}$ are shown for each module  $R_i=\bigoplus_j L_2(D_6,\lambda_j)\otimes M_7(h_{ij})$.}
\def\arraystretch{1.1}
    \centering
    \begin{tabular}{c|c|c|c|c|c|c|c|c|c|ccc} \hline
$\omega_i$ & $R_i$   &  $0$  &  $ 2w_6$    &   $ 2w_1$   & $2w_5$   &   $ w_2$    &  $w_4$  &  $w_6 $  &  $ w_1+w_5$  & $\mu_i$
 \\ \hline    
$0$  &$R_{0}$ &    $  -1/8  $   &   $  43/8  $   &    $  39/8  $   &     $  3/8  $  &    $  25/24  $   &    $  37/24  $   &     $  3/16  $  &    $  43/16  $  & $0 $  \\ 
$ 2w_6$    &$R_{1}$ &      $  3/2  $   &   $ 0  $   &    $ 15/2   $   &     $  7  $  &    $  8/3  $   &    $  13/6  $   &     $   13/16 $  &    $  69/16  $  &    $  13/8  $     \\ 
$ 2w_1$   &$R_{2}$ &     $  0  $   &   $  3/2  $   &    $  7 $   &     $  15/2  $  &    $ 13/6   $   &    $  8/3  $   &     $  69/16  $  &    $  13/16  $  &    $  9/8  $  \\ 
$2w_5$   &$R_{3}$ &     $   43/8 $   &   $  -1/8  $   &    $   3/8 $   &     $  39/8  $  &    $  37/24  $   &    $ 25/24   $   &     $  43/16  $  &  $  3/16  $  &      $ 3/2   $  \\ 
$ w_2$    &$R_{4}$ &   $   3/4 $   &   $   9/4 $   &    $   7/4 $   &     $ 5/4   $  &    $  -1/12  $   &    $  5/12  $   &     $  1/16  $  &$ 9/16   $  &    $  7/8  $  \\ 
$w_4$  &$R_{5}$ &    $  9/4  $   &   $ 3/4   $   &    $  5/4  $   &     $   7/4 $  &    $   5/12 $   &    $  -1/12  $   &     $   9/16 $  &$   1/16 $  &    $  11/8  $ \\ 
$w_6 $  &$R_{6}$ &   $  5/8  $   &   $  25/8  $   &    $   29/8 $   &     $  1/8  $  &    $  19/24  $   &    $ 7/24   $   &     $    -1/16$  &  $  23/16  $  &  $  3/4  $ \\ 
 $ w_1+w_5$  &$R_{7}$ &    $ 25/8   $   &   $ 5/8   $   & $   1/8 $  &   $  29/8  $   &     $  7/24  $  &    $  19/24  $   &    $  23/16  $   &     $  -1/16  $  &    $   5/4 $   \\ \hline
     \end{tabular}
    \label{tab:WE7Ramondirrbosonic}
\end{table}

\begin{table}[ht]
\caption{All Ramond-twisted irreducible modules of $\cW_{-2}(E_7,f_\theta)$ that form the R sector. The $h_{ij}$ are shown for each module $R_i=\bigoplus_j L_2(D_6,\lambda_j)\otimes M_7(h_{ij})$.}
\def\arraystretch{1.1}
    \centering
    \begin{tabular}{c|c|c|c|c|c|c|cccccc} \hline
$\omega_i$ & $R_i$ & $w_1 $  & $w_5+w_6 $    &  $ w_3  $  &  $w_5  $    &   $w_1+w_6  $      & $\mu$     
 \\ \hline    
$w_1 $  &$R_{8}$ &   $  -13/192   $  &    $  179/192   $  &    $ 81/64    $   &   $   45/64  $   &    $   13/64  $    &    $   33/64  $      \\ 
$w_5+w_6 $    & $R_{9}$ &    $  179/192   $  &    $  -13/192   $  &    $ 81/64    $   &   $   45/64  $   &    $   13/64  $   &    $ 97/64    $       \\ 
$ w_3  $  &$R_{10}$ &     $  311/192   $  &$  311/192   $  &    $   -3/64,381/64  $  &    $   25/64,217/64  $      &    $   57/64,249/64  $     &       $  77/64   $       \\ 
$w_5  $    &$R_{11}$ &      $   23/192  $  &    $  23/192   $  &    $ 29/64 ,157/64   $   &   $  -7/64,57/64   $   &    $  25/64,89/64   $        &       $  45/64   $        \\ 
$w_1+w_4  $      & $R_{12}$ &     $  131/192   $  &    $ 131/192    $  &    $ 1/64,257/64    $   &   $ 29/64,157/64    $   &    $  -3/64,125/64   $      &       $  81/64   $       \\  \hline
     \end{tabular}
    \label{tab:WE7Ramondirrfermionic}
\end{table}

\subsection{Irreducible modules of $\cW_{-2}(E_7,f_\theta)$}
To recover the irreducible modules from the Ramond-twisted ones of $\cW_{-2}(E_7,f_\theta)$, we need to use the simple current of $L_2(D_6)$. The affine VOA 
$L_2(D_6)$ has four simple currents, and we use $L_2(D_6,2w_6,3/2)$ which  induces a $\ZZ_2$ automorphism for the irreducible modules of $L_2(D_6)$. Besides the exchange between the vacuum and $L_2(D_6,2w_6,3/2)$, the simple current induces the following exchanges
\begin{align}\nonumber
L_2(D_6,2w_1,1)&\leftrightarrow L_2(D_6,2w_5,3/2),\\ \nonumber   
L_2(D_6,w_2,5/6)&\leftrightarrow L_2(D_6,  w_4,4/3) ,\\ \nonumber
L_2(D_6,w_5,11/16)& \leftrightarrow L_2(D_6,w_1+w_6,19/16),\\ \nonumber  L_2(D_6,w_1,11/24 )&\leftrightarrow L_2(D_6,w_5+w_6, 35/24) .
\end{align}
The rest three irreducible modules fusion to themselves
\begin{align}\nonumber
   L_2(D_6,w_3,9/8),\quad L_2(D_6,w_6,11/16),\quad  L_2(D_6,w_1+w_5,19/16).
\end{align}

By this $\ZZ_2$ automorphism for the $L_2(D_6)$ irreducible modules, we can easily obtain all 13 irreducible modules $N_0,N_1,\dots,N_{12}$ of $\cW_{-2}(E_7,f_\theta)$ from the Ramond-twisted ones in Tables \ref{tab:WE7Ramondirrbosonic} and \ref{tab:WE7Ramondirrfermionic}. For example, the $R_1$ gives $N_0\cong \cW_{-2}(E_7,f_\theta)$ in \eqref{isoE7}. We show one more example here. From the $R_{8}$ in Table \ref{tab:WE7Ramondirrfermionic}, we obtain the following irreducible module of $\cW_{-2}(E_7,f_\theta)$:
\begin{align}\nonumber
    N_8\cong&\, L_2(D_6, w_1)\otimes M_7 \bigg(\frac{ 179}{ 192}\bigg)\oplus L_2(D_6, w_5+w_6)\otimes M_7 \bigg(\!-\frac{ 13}{ 192}\bigg)\oplus L_2(D_6, w_3)\otimes M_7 \bigg(\frac{ 81}{ 64}\bigg)\\ \nonumber
    &\oplus L_2(D_6, w_5)\otimes M_7 \bigg(\frac{ 45}{ 64}\bigg)\oplus L_2(D_6, w_1+w_6)\otimes M_7 \bigg(\frac{ 13}{ 64}\bigg).
\end{align}
In fact, $R_8\cong N_8\oplus \CC^{12}$ which can be deduced by comparing the character $\chi(R_8)-\chi(N_8)=12$.

Finally, we comment on the $\ZZ_2=\{1,\sigma\}$ automorphism group of $\cW_{-2}(E_7,f_\theta)$. In \cite{Lee:2024fxa}, $w_1,w_3$ and $w_6$ of $D_6$ are called fermionic fundamental weights. A highest weight of $D_6$ irreducible is called bosonic/fermionic if it contains an even/odd number of fermionic fundamental weights. 
Then in Table \ref{tab:WE7Ramondirrbosonic}, the first six columns from $0$ to $w_4$ are bosonic weights while  the last two columns $w_6$ and $w_1+w_5$ are fermionic weights of $D_6$. In Table \ref{tab:WE7Ramondirrfermionic}, the first three columns from $w_1$ to $w_3$ are fermionic weights while  the last two columns $w_5$ and $w_1+w_6$ are bosonic weights of $D_6$. According to the previous discussion, the even and odd parts of $\cW_{-2}(E_7,f_\theta)$ are contributed by the even and odd irreducible modules of of $L_2(D_6)$ respectively. This suggest that $\sigma(b)=b$ and $\sigma(f)=-f$ for any $b $ in $L_2(D_6)\otimes M_7$ modules in the first six columns and $f$ in $L_2(D_6)\otimes M_7$ modules in the last two columns in Table \ref{tab:WE7Ramondirrbosonic}. Similar for all other irreducible modules $N_i$.
 
\subsection{Coset B}
$L_2(\Dh)$ is an intermediate vertex subalgebra of $L_2(E_7)$. We find the coset construction
 \begin{align}\label{cosetED}
     \frac{L_2(E_7)}{ L_2(\Dh)}=\mathrm{Vir}^{\rm eff}_{16,5}.
 \end{align}
The minimal model $\mathrm{Vir}_{16,5}$ has central charge $-{323}/{40}$ and effective central charge $ {37}/{40}$. It has $30$ irreducible modules. We compute their characters as $q$-series to sufficient high orders. We then check the character relations for the coset for every irreducible module of $L_2(E_7)$. We collect all module decompositions of $L_2(E_7)$ in the coset construction \eqref{cosetED} in Table \ref{tab:CosetED}. We use $L_2(E_7,\omega_i,\mu_i)$ to denote the $L_2(E_7)$ irreducible module with $E_7$ highest-weight  $\omega_i$ and conformal dimension $\mu_i$. The highest weights $2w_6,w_1,w_5$ give $E_7$ irreducible modules $\bf 1463, 133, 1539 $, while $w_6,w_7$ give $\bf 56, 912$. The two tabulars in Table \ref{tab:CosetED} present the bosonic and fermionic  $E_7$ irreducible modules respectively. We checked that all $L_2(E_7)$ module decompositions of \eqref{cosetED} are consistent with the module decompositions for $\Dh\subset E_7$ in \cite{Lee:2024fxa}.

\begin{table}[ht]
\caption{Coset construction $L_2(E_7,\omega_i,\mu_i)=\bigoplus_j L_2(\Dh,\lambda_j)\otimes \mathrm{Vir}^{\rm eff}_{16,5}(h_{ij})$.}
\def\arraystretch{1.1}
    \centering
    \begin{tabular}{c|c|c|c|c|c|c|c|c|cccc} \hline
$\omega_i$     &  $0$  &  $ 2w_5$    &   $ w_6$   & $ w_1+w_5 $   &   $ w_2 $    &  $ w_4 $  &  $ 2w_1  $  &  $ 2w_6 $  & $\mu_i$
 \\ \hline    
$0$    &    $ 0    $   &   $  15/2   $   &    $  1/4   $   &     $   19/4 $  &    $ 9/8    $   &    $ 21/8   $   &     $    87/8 $  &    $    3/8 $  & $0 $  \\ 
$2w_6 $    &    $  15/2   $   &   $   0  $   &    $  19/4   $   &     $   1/4 $  &    $   21/8  $   &    $  9/8  $   &     $  3/8   $  &    $   87/8  $  & $3/2 $  \\
$ w_1$    &    $   9/10  $   &   $  17/5   $   &    $  3/20   $   &     $  33/20  $  &    $   1/40  $   &    $   21/40 $   &     $  231/40   $  &    $   91/40  $  & $9/10 $  \\
$w_5 $    &    $  17/5   $   &   $  9/10   $   &    $   33/20  $   &     $  3/20  $  &    $ 21/40    $   &    $  1/40  $   &     $  91/40   $  &    $  231/40   $  & $ 7/5$  \\
\hline
     \end{tabular}
     \\[+1mm]
      \begin{tabular}{c|c|c|c|c|cccc} \hline
$\omega_i$     &  $w_1 ,  w_5+w_6$    &   $w_5$   & $ w_3  $   &   $ w_1+w_6 $       & $\mu_i$
 \\ \hline    
$w_6 $    &    $  63/320   $      &    $ 3/320,323/320  $   &     $  483/320,1443/320   $  &    $  143/320,783/320   $   &    $ 57/80 $  \\ 
$ w_7$    &    $    115/64     $   &    $ 39/64,231/64  $   &     $  7/64,583/64   $  &    $  3/64, 387/64   $   &     $ 21/16 $  \\
%$ $    &    $     $   &   $      $   &    $   $   &     $     $  &    $     $   &    $     $   &     $     $  &    $     $  & $  $  \\
\hline
     \end{tabular}
    \label{tab:CosetED}
\end{table}

As both $L_2(E_7)$ and $L_2(\Dh)$ allow fermionization,  a super version of  coset \eqref{cosetED} has been found in \cite[Equation (4.114)]{Lee:2024fxa} as 
$
{\cF L_2(E_7)}/{\cF L_2(\Dh)}=F^{-1}\otimes (\mathrm{Vir}_{80,2}^{N=1})^{\rm e}_{\rm eff}.
$
Here $F$ is the free fermion SVOA with central charge $1/2$. The extension of $\mathrm{Vir}_{80,2}^{N=1}$ is constructed in \cite[Equation (4.124)]{Lee:2024fxa} which has effective central charge $57/40$. We remark that the coset and super coset are often independent. For example from \eqref{cosDh}, one might expect ${\cF L_2(\Dh)}/{\cF L_2(D_6)}$ gives  just $ (\mathrm{Vir}_{16,6}^{N=1} )_{\rm eff}$. However, we checked this is not true. 

\section{$\cW_{-1}(E_6,f_\theta)$}\label{sec:E6}
The $\cW$-algebra $\cW_{-1}(E_6,f_\theta)$ has central charge $c={76}/{11}$. The affine VOA $L_2(A_5)$ has central charge $c={35}/{4}$.  
The intermediate vertex subalgebra $L_2(\Ah)\subset L_2(E_6)$ has central charge $c= {112}/{11}.$   We find the following coset construction
\begin{align}\label{cosE6}
   \frac{\cW_{-1}(E_6,f_\theta)}{ L_2(A_5)} =( \cB\mathrm{Vir}_{22,8}^{N=1})^{\rm e}
\end{align}
and 
\begin{align}\label{cosAh}
    \frac{L_2(\Ah)}{ L_2(A_5) }=( \cB\mathrm{Vir}_{22,8}^{N=1})^{\rm e}_{\rm eff}.
\end{align}
To verify these cosets, we first find the characters of $L_2(\Ah)$, which is achieved by a rather complicated construction from Hecke operator.  We then use these cosets to determine all irreducible modules  and  Ramond-twisted irreducible modules  of $\cW_{-1}(E_6,f_\theta)$.

\subsection{Hecke operator}
It was conjectured in \cite{Lee:2024fxa} that the characters of $L_2(\Ah)$ may be realized as a $\T_7$ Hecke image of some RCFT with central charge $c= {16}/{11}$. We now construct this RCFT as a non-diagonal modular invariant of the bosonization of the $N=1$ supersymmetric minimal model $\mathrm{Vir}_{132,2}^{N=1}$. 

From \eqref{ceff}, it is easy to check that $\mathrm{Vir}_{132,2}^{N=1}$ indeed has effective central charge $16/11$. It has 33 Neveu-Schwarz irreducible modules with effective NS dimensions
\begin{align}\label{phi1322}
&\bigg\{ 0,\frac{1}{66},\frac{1}{22},\frac{1}{11},\frac{5}{33},\frac{5}{22},\frac{7}{22},\frac{14}{33},\frac{6}{11},\frac{15}{22},\frac{5}{6},1,\frac{13}{11},\frac{91}{66},\frac{35}{22},\frac{20}{11},\frac{68}{33},\frac{51}{22},\\ \nonumber
 &\ \ \frac{57}{22},\frac{95}{33},\frac{35}{11},\frac{7}{2},\frac{23}{6},\frac{46}{11},\frac{50}{11},\frac{325}{66},\frac{117}{22},\frac{63}{11},\frac{203}{33},\frac{145}{22},\frac{155}{22},\frac{248}{33},8  \bigg\} ,
\end{align}
and 33 Ramond irreducible modules with effective R dimensions
\begin{align}\label{psi1322}
 &\bigg\{\frac{2}{33},\frac{3}{44},\frac{1}{11},\frac{17}{132},\frac{2}{11},\frac{1}{4},\frac{1}{3},\frac{19}{44},\frac{6}{11},\frac{89}{132},\frac{9}{11},\frac{43}{44},\frac{38}{33},\frac{59}{44},\frac{17}{11},\frac{233}{132},2,\frac{9}{4},\\ \nonumber
 &\ \ \frac{83}{33},\frac{123}{44},\frac{34}{11},\frac{449}{132},\frac{41}{11},\frac{179}{44},\frac{146}{33},\frac{211}{44},\frac{57}{11},\frac{67}{12},6,\frac{283}{44},\frac{227}{33},\frac{323}{44},\frac{86}{11}   \bigg\}  .
\end{align}
The bosonization gives a rational VOA with $33\times 3+1=100$ irreducible modules with effective conformal dimensions from (i) $\phi$ sector: all dimensions in \eqref{phi1322}, (ii) $\varphi$ sector: all dimensions in \eqref{phi1322} added by $1/2$ except the last one added by $3/2$, (iii) $\psi$ sector: all dimensions in \eqref{psi1322} with one more dimension $35/32$. The last one is due to the presence of zero exponent of the R sector as in the $\cB\mathrm{Vir}_{16,6}^{N=1}$ case.

With all 100 irreducible modules and characters of $\cB\mathrm{Vir}_{132,2}^{N=1}$ at hand, we construct the following non-diagonal modular invariant with 12 characters (up to degeneracy)
\begin{align}\nonumber
    \chi_0^{\rm e}&=\phi_0+\phi_1+\psi_2-\varphi_4+\psi_6+\phi_8,\\ \nonumber
    \chi_{\frac{2}{33}}^{\rm e}&=  \psi_{\frac{2}{33}}+\psi_{\frac{35}{33}}+2\phi_{\frac{68}{33}},\\\nonumber
     \chi_{\frac{5}{33}}^{\rm e}&=  \phi_{\frac{5}{33}}+\psi_{\frac{38}{33}}+\phi_{\frac{203}{33}},\\\nonumber
     \chi_{\frac{2 }{11}}^{\rm e}&= \psi_{\frac{2}{11}}+ \phi_{\frac{13}{11}} - \varphi_{\frac{13}{11}} +\phi_{\frac{35}{11}}  +\phi_{\frac{46}{11}}   - \psi_{\frac{57}{11}} ,\\\nonumber
     \chi_{\frac13}^{\rm e}&=\psi_{\frac13}+\varphi_{\frac43}+\varphi_{\frac{13}{3}}, \\\nonumber
     \chi_{\frac{14 }{33}}^{\rm e}&= \phi_{\frac{14}{33}}+\psi_{\frac{146}{33}}-\varphi_{\frac{179}{33}}  ,\\\nonumber
     \chi_{\frac{17 }{33}}^{\rm e}&= \varphi_{\frac{17}{33}} + \psi_{\frac{83}{33}}  - \phi_{\frac{248}{33}},\\\nonumber
     \chi_{\frac{6 }{11}}^{\rm e}&=  \phi_{\frac{6}{11}}-\varphi_{\frac{6}{11}}+\psi_{\frac{6}{11}}-\psi_{\frac{17}{11}}+\phi_{\frac{50}{11}}-\varphi_{\frac{83}{11}} ,\\\nonumber
     \chi_{\frac{ 8}{11}}^{\rm e}&=  \varphi_{\frac{8}{11}} +\psi_{\frac{41}{11}}-\phi_{\frac{63}{11}},\\\nonumber
     \chi_{\frac{ 9}{11}}^{\rm e}&= \varphi_{\frac{9}{11}}+\psi_{\frac{9}{11}}  -\phi_{\frac{20}{11}}+\varphi_{\frac{31}{11}} +\varphi_{\frac{64}{11}} -\psi_{\frac{86}{11}} ,\\\nonumber
     \chi_{\frac{12}{11}}^{\rm e}&=\psi_{\frac{1}{11}} -\phi_{\frac{1}{11}}+\varphi_{\frac{23}{11}}+\varphi_{\frac{34}{11}}-\psi_{\frac{34}{11}}+\varphi_{\frac{78}{11}}  ,\\\nonumber
     \chi_{\frac{ 62}{33}}^{\rm e}&=\varphi_{\frac{ 62}{33}}  - \phi_{\frac{ 95}{33}}   + \psi_{\frac{ 227}{33}}  .
\end{align}
We compute the $100\times 100$ $S$-matrix of $\cB\mathrm{Vir}_{132,2}^{N=1}$ and verify that the following combination is modular invariant under $SL(2,\ZZ)$:
\begin{align}
Z^{\rm e}=  &\  |\chi_0^{\rm e}|^2+\big|\chi_{\frac{2}{35}}^{\rm e}\big|^2 +2\big|\chi_{\frac{5}{35}}^{\rm e}\big|^2+\big|\chi_{\frac{2}{11}}^{\rm e}\big|^2 +2\big|\chi_{\frac{1}{3}}^{\rm e}\big|^2+2\big|\chi_{\frac{14}{35}}^{\rm e}\big|^2\\ \nonumber
&\,+2\big|\chi_{\frac{17}{35}}^{\rm e}\big|^2+\big|\chi_{\frac{6}{11}}^{\rm e}\big|^2 +2\big|\chi_{\frac{8}{11}}^{\rm e}\big|^2 +\big|\chi_{\frac{9}{11}}^{\rm e}\big|^2+\big|\chi_{\frac{12}{11}}^{\rm e}\big|^2 +2\big|\chi_{\frac{62}{33}}^{\rm e}\big|^2 .
\end{align}
This is a 12-character modular invariant with 6 degeneracy two. 
Equivalently, one can regard this as a 13-character modular invariant with 8 degeneracy two, thanks to the relation
\begin{align}\nonumber
    \big|\chi_{\frac{2}{35}}^{\rm e}\big|^2+1=2\big|\psi_{\frac{2}{33}}+\phi_{\frac{68}{33}}\big|^2+2\big|\psi_{\frac{35}{33}}+\phi_{\frac{68}{33}}\big|^2.
\end{align}
Then the 13 characters satisfy a MLDE with $d=13$ and index $l=36$.

By Hecke operator $\T_7$ on the characters of the above non-diagonal modular invariant, we obtain a vector valued modular form with 13 components as follows
\begin{align}\nonumber
 \chi_0 &= q^{-\frac{14}{33}}(1+56 q+1652 q^2+19936 q^3+163226 q^4+1036224 q^5 +\dots )  ,\\\nonumber
  \chi_{\frac{14}{33}} &= 6+336 q+6090 q^2+60816 q^3+441336 q^4+2583672 q^5+\dots  ,\\ \nonumber
  \chi_{\frac{20}{33}} &=  q^{\frac{2}{11}}( 21+1176 q+18312 q^2+170736 q^3+1183875 q^4+ \dots ) ,\\ \nonumber
  \chi_{\frac{8}{11}} &= q^{\frac{10}{33}}( 21+1176 q+16716 q^2+149352 q^3+1006488 q^4 +\dots )  ,\\ \nonumber
  \chi_{\frac{9}{11}} &=  q^{\frac{13}{33}}( 56+1960 q+26656 q^2+230336 q^3+1522192 q^4+ \dots ) ,\\ \nonumber
  \chi_{\frac{32}{33}} &=  q^{\frac{6}{11}}( 21+504 q+6420 q^2+52584 q^3+336861 q^4+ \dots ) ,\\ \label{chiA0}
  \chi_{\frac{35}{33}} &=  q^{\frac{7}{11}}(120+3045 q+36120 q^2+289380 q^3+1816416 q^4+ \dots ) ,\\\nonumber
  \chi_{\frac{12}{11}} &=  q^{\frac{2}{3}}(105+2520 q+29526 q^2+234304 q^3+1461935 q^4+ \dots ) ,\\\nonumber
  \chi_{\frac{38}{33}} &=  q^{\frac{8}{11}}(210+5040 q+56721 q^2+443352 q^3+2731260 q^4+ \dots ) ,\\\nonumber
  \chi_{\frac{14}{11}} &=  q^{\frac{28}{33}}( 384+7896 q+84420 q^2+637056 q^3+3835860 q^4+ \dots ) ,\\\nonumber
  \chi_{\frac{4}{3}} &=  q^{\frac{10}{11}}( 210+3984 q+41706 q^2+309120 q^3+1841406 q^4+ \dots ) ,\\\nonumber
  \chi_{\frac{47}{33}} &=  336+6090 q+60816 q^2+441336 q^3+2583672 q^4+\dots ,\\ \nonumber
  \chi_{\frac{18}{11}} &=  q^{\frac{40}{33}}( 196+2912 q+26999 q^2+185416 q^3+1047620 q^4+ \dots ) .
\end{align}
We observe that these exactly describe the 13 characters of $L_2(\Ah)$. The leading Fourier coefficients give exactly the irreducible representation dimensions of $\Ah$ predicted in \cite[Table 10]{Lee:2024fxa}. All conformal dimensions in \eqref{chiA0} are also consistent with the quadratic Casimir invariants of $\Ah$ found in \cite[Table 10]{Lee:2024fxa}. The degeneracy also completely agrees with those predicted in \cite{Lee:2024fxa}. The conformal dimensions and degeneracy are summarized in the following subsection in \eqref{eq:Ahlevel2}. Note $\chi_{\frac{47}{33}}-\chi_{\frac{14}{33}}=6$, similar to the $L_2(\Dh)$ case.

\subsection{Coset A}
The $A_5,\Ah$ and $E_6$ all have a $\ZZ_3$ center, from which $L_2(A_5)$, $L_2(\Ah)$, $L_2(E_6) $ and $\cW_{-1}(E_6,f_\theta)$ inherit a $\ZZ_3$ automorphism. This $\ZZ_3$ automorphism is useful for verifying the character relations for coset \eqref{cosAh}. 
The affine VOA $L_2(A_5)$ has 21 irreducible modules with conformal dimensions
\begin{align}\label{eq:Ah2conformald}
  0,\left(\frac{35}{96}\right)_2, \left(\frac{7}{12}\right)_2,\frac{21}{32},\frac{3}{4},\left(\frac{5}{6}\right)_2,\left(\frac{95}{96}\right)_2,\left(\frac{33}{32}\right)_2,\left(\frac{13}{12}\right)_2,\frac{5}{4},\left(\frac{4}{3}\right)_2,\left(\frac{131}{96}\right)_2,\frac{3}{2}  .
\end{align}
The $\mathbb{Z}_3$ automorphism organizes the 21 irreducible modules into 7 orbits
\begin{align} \label{Z3A5h2}
\big(\phi_0,\phi_{\frac43},\bar\phi_{\frac43}\big),\big(\phi_{\frac{7}{12}},&\,\bar\phi_{\frac{7}{12}},\phi_{\frac{5}{4}}\big),\big(\phi_{\frac{21}{32}},\phi_{\frac{95}{96}},\bar\phi_{\frac{95}{96}}\big),\big(\phi_{\frac{3}{4}},\phi_{\frac{13}{12}},\bar\phi_{\frac{13}{12}}\big),\big(\phi_{\frac{5}{6}},\bar\phi_{\frac{5}{6}},\phi_{\frac{3}{2}}\big),\\ \nonumber
&\big(\phi_{\frac{35}{96}},\bar\phi_{\frac{35}{96}},\phi_{\frac{33}{32}}\big),\big(\phi_{\frac{33}{32}},\phi_{\frac{131}{96}},\bar\phi_{\frac{131}{96}}\big).
\end{align}
In each orbit, the differences between conformal dimensions are multiples of $1/3$. Denote $\mathbb{Z}_3=\{1,\rho,\rho^2\}$. Then each orbit can be written as $(\phi_h, \rho \phi_h,\rho^2 \phi_h)$. The $\rho \phi_h$ and $\rho^2 \phi_h$ are conjugate to each other due to the outer automorphism $\ZZ_2$ of $A_5$.

As predicted in \cite{Lee:2024fxa} and confirmed by the Hecke operator computation, the intermediate vertex subalgebra $L_2(\Ah)$ has also 21 irreducible modules with the following conformal dimensions
\begin{align}\label{eq:Ahlevel2}
0,\left(\frac{14}{33}\right)_2,\left(\frac{20}{33}\right)_2,\frac{8}{11},\frac{9}{11},\left(\frac{32}{33}\right)_2,\left(\frac{35}{33}\right)_2,\left(\frac{12}{11}\right)_2,\left(\frac{38}{33}\right)_2,\frac{14}{11},\left(\frac{4}{3}\right)_2,\left(\frac{47}{33}\right)_2 ,\frac{18}{11} .
\end{align}
Similar with $L_2(A_5)$, 
the $\mathbb{Z}_3$ automorphism organizes these 21 irreducible modules into the following 7 orbits
\begin{align} 
\big(\phi_0,\phi_{\frac43},\bar\phi_{\frac43}\big),\big(\phi_{\frac{20}{33}},&\,\bar\phi_{\frac{20}{33}},\phi_{\frac{14}{11}}\big),\big(\phi_{\frac{8}{11}},\phi_{\frac{35}{33}},\bar\phi_{\frac{35}{33}}\big),\big(\phi_{\frac{9}{11}},\phi_{\frac{38}{33}},\bar\phi_{\frac{38}{33}}\big),\big(\phi_{\frac{32}{33}},\bar\phi_{\frac{32}{33}},\phi_{\frac{18}{11}}\big),\\ \nonumber
&\big(\phi_{\frac{14}{33}},\bar\phi_{\frac{14}{33}},\phi_{\frac{12}{11}}\big),\big(\phi_{\frac{12}{11}},\phi_{\frac{47}{33}},\bar\phi_{\frac{47}{33}}\big).
\end{align}

The $N=1$ minimal model $\mathrm{Vir}_{22,8}^{N=1}$ has central charge $- {81}/{44}$ and effective central charge ${63}/{44}$. 
It has 37 irreducible Neveu-Schwarz modules and 37 irreducible Ramond modules. We find the bosonization $\cB\mathrm{Vir}_{22,8}^{N=1}$ has in total $37\times 3+1=112$ irreducible modules. To construct the coset, we first construct an extension of $\cB\mathrm{Vir}_{22,8}^{N=1}$, i.e., a non-diagonal modular invariant.  
We define the following 10 extensions from the $\phi$ sector of $\cB\mathrm{Vir}_{22,8}^{N=1}$:
\begin{align}\nonumber
  &\chi_{-\frac{3}{22}}=\phi_{-\frac{3}{22}}+\phi_{\frac{261}{22}},\ \chi_{-\frac{5}{44}}= \phi_{-\frac{5}{44}}+\phi_{\frac{39}{44}}, \ \chi_{-\frac{3}{44}}= \phi_{-\frac{3}{44}}+\phi_{\frac{85}{44}},\ \chi_0=\phi_0+\phi_{15},\ \chi_{\frac{1}{11}}=\phi_{\frac{1}{11}}+\phi_{\frac{100}{11}},\\ \nonumber
  &\chi_{\frac{15}{44}}= \phi_{\frac{15}{44}}+\phi_{\frac{147}{44}},\ \chi_{\frac{15}{22}}=\phi_{\frac{15}{22}}+\phi_{\frac{147}{22}},\ \chi_{\frac{19}{44}}= \phi_{\frac{19}{44}}+\phi_{\frac{225}{44}},\ \chi_{\frac{18}{11}}=\phi_{\frac{18}{11}}+\phi_{\frac{51}{11}},\ \chi_{\frac{9}{4}}= \phi_{\frac{9}{4}}+\phi_{\frac{29}{4}},
\end{align}
and the following 10 extensions from the $\varphi$ sector:
\begin{align}\nonumber
   & \chi_{\frac{4}{11}}=\varphi_{\frac{4}{11}}+\varphi_{\frac{136}{11}},\ \chi_{\frac{17}{44}}=\varphi_{\frac{17}{44}}+\varphi_{\frac{61}{44}},\ 
    \chi_{\frac{19}{44}}=\varphi_{\frac{19}{44}}+\varphi_{\frac{107}{44}},\ \chi_{\frac{13}{22}}=\varphi_{\frac{13}{22}}+\varphi_{\frac{211}{22}},\ \chi_{\frac{37}{44}}=\varphi_{\frac{37}{44}}+\varphi_{\frac{169}{44}},\\ \nonumber
 &   \chi_{\frac{13}{11}}=\varphi_{\frac{13}{11}}+\varphi_{\frac{79}{11}},\ \chi_{\frac{3}{2}}=\varphi_{\frac{3}{2}}+\varphi_{\frac{31}{2}},\ \chi_{\frac{71}{44}}=\varphi_{\frac{71}{44}}+\varphi_{\frac{247}{44}},\ \chi_{\frac{47}{22}}=\varphi_{\frac{47}{22}}+\varphi_{\frac{113}{22}},\ \chi_{\frac{11}{4}}=\varphi_{\frac{11}{4}}+\varphi_{\frac{31}{4}},
\end{align}
and the following 5 extensions from the $\psi$ sector:
\begin{align}\nonumber
\chi_{-\frac{23}{352}}&= \psi_{-\frac{23}{352}}+\psi_{\frac{2089}{352}},\ \chi_{\frac{9}{352}}= \psi_{\frac{9}{352}}+\psi_{\frac{1417}{352}} ,\ \chi_{\frac{73}{352}}= \psi_{\frac{73}{352}}+\psi_{\frac{2889}{352}},\\ \nonumber
&\quad\quad  \chi_{\frac{169}{352}}= \psi_{\frac{169}{352}}+\psi_{\frac{873}{352}},\ \chi_{\frac{27}{32}}= \psi_{\frac{27}{32}}+\psi_{\frac{347}{32}}   .
\end{align}
We also take $\chi_{h_i}=\psi_{h_i}$ from the $\psi$ sector for the $h_i$ appearing in the following modular invariant \eqref{eq:e228} with degeneracy two.

We explicitly compute the $112\times 112$ $S$-matrix of $\cB\mathrm{Vir}^{N=1}_{22,8}$ and verify that the following combination is modular invariant under $SL(2,\ZZ)$:
\begin{align}\label{eq:e228}
    Z=&\ \big|\chi_{-\frac{3}{22}}\big|^2+\big|\chi_{-\frac{5}{44}}\big|^2+\big|\chi_{-\frac{3}{44}}\big|^2+\big|\chi_{-\frac{23}{352}}\big|^2+\big|\chi_{0}\big|^2+\big|\chi_{\frac{9}{352}}\big|^2+\big|\chi_{\frac{1}{11}}\big|^2+\big|\chi_{\frac{73}{352}}\big|^2\\ \nonumber
    &+\big|\chi_{\frac{15}{44}}\big|^2+\big|\chi_{\frac{4}{11}}\big|^2+\big|\chi_{\frac{17}{44}}\big|^2+\big|\chi_{\frac{19}{44}}\big|^2+\big|\chi_{\frac{169}{352}}\big|^2+\big|\chi_{\frac{13}{22}}\big|^2+\big|\chi_{\frac{15}{22}}\big|^2+\big|\chi_{\frac{37}{44}}\big|^2\\ \nonumber
    &+\big|\chi_{\frac{27}{32}}\big|^2+\big|\chi_{\frac{49}{44}}\big|^2+\big|\chi_{\frac{13}{11}}\big|^2+\big|\chi_{\frac{3}{2}}\big|^2+\big|\chi_{\frac{71}{44}}\big|^2+\big|\chi_{\frac{18}{11}}\big|^2+\big|\chi_{\frac{47}{22}}\big|^2+\big|\chi_{\frac{9}{4}}\big|^2\\ \nonumber
    &+\big|\chi_{\frac{11}{4}}\big|^2+2\big|\chi_{-\frac{27}{352}}\big|^2+2\big|\chi_{\frac{37}{352}}\big|^2+2\big|\chi_{\frac{9}{44}}\big|^2+2\big|\chi_{\frac{229}{352}}\big|^2+2\big|\chi_{\frac{31}{44}}\big|^2+2\big|\chi_{\frac{325}{352}}\big|^2\\ \nonumber
    &+2\big|\chi_{\frac{457}{352}}\big|^2+2\big|\chi_{\frac{549}{352}}\big|^2+2\big|\chi_{\frac{997}{352}}\big|^2+2\big|\chi_{\frac{65}{22}}\big|^2+2\big|\chi_{\frac{38}{11}}\big|^2+2\big|\chi_{\frac{143}{32}}\big|^2.
\end{align}
This shows that $\cB\mathrm{Vir}^{N=1}_{22,8}$ has a non-diagonal modular invariant with 49 irreducible modules and 37 distinct characters. We denote this extension as $(\cB\mathrm{Vir}^{N=1}_{22,8})^{\rm e}$.

Now we verify the coset \eqref{cosAh} by checking the character relations to sufficiently high $q$ degrees. For example, 
for the vacuum character of $L_2(\Ah)$, i.e., the $\chi_0$ in \eqref{chiA0}, we find 
\begin{align}\label{chiAh0}
 \chi_0=\chi_0^{\cB} \chi^{A}_0 +\chi_{\frac{5}{4}}^{\cB} \chi^{A}_{\frac34}+\chi_{\frac74}^{\cB} \chi^{A}_{\frac54}+\chi_{\frac12}^{\cB} \chi^{A}_{\frac32} +\chi_{\frac{11}{32}}^{\cB}\chi^{A}_{\frac{21}{32}}+2 \chi_{\frac{95}{32}}^{\cB}\chi^{A}_{\frac{33}{32}}.
\end{align}
Here $ A$ is short for $ {L_2(A_5)}$ and $ {\cB}$ is short for effective $(\cB\mathrm{Vir}^{N=1}_{22,8})^{\rm e}$. The weight-1 space are contributed from the first and fifth terms in the right hand side of \eqref{chiAh0}, which gives exactly the module decomposition $\bf 56=\bf 35+20+1$ for $A_5\subset \Ah$ in \cite{Lee:2024fxa}.   
Now consider the $\rho$ image of the vacuum in $\ZZ_3$ automorphism, i.e., the $\chi_{4/3}$ in \eqref{chiA0}, we find
\begin{align}
\chi_{\frac{4}{3}}= \chi_0^{\cB}  \chi^{A}_{\frac43} +\chi_{\frac54}^{\cB}\chi^{A}_{\frac{13}{12}}+\chi_{\frac74}^{\cB}\chi^{A}_{\frac{7}{12}}+\chi_{\frac12} ^{\cB}\chi^{A}_{\frac56}+\chi_{\frac{11}{32}}^{\cB} \chi^{A}_{\frac{95}{96}}+\chi_{\frac{95}{32}}^{\cB}\big(\chi^{A}_{\frac{35}{96}}+\chi^{A}_{\frac{131}{96}}\big).
\end{align}
The weight-0 space  gives precisely the $A_5\subset \Ah$ module decomposition $\bf 210'=\bf  105' + \overline{84} + 21$  in \cite{Lee:2024fxa}. 
Comparing with \eqref{chiAh0}, 
we observe that the $\cB$ modules do not change, while the ${A} $ modules change to its $\rho$ image. This persists to be true for all other six $\ZZ_3$ orbits.

By a large amount of computation, we find and verify the character relations of coset \eqref{cosAh} for all 21 characters of $L_2(\Ah)$ in \eqref{chiA0}. Then by carefully comparing with the module decompositions under $A_5\subset \Ah$ in \cite{Lee:2024fxa}, we further upgrade the character relations to module decompositions for $L_2(\Ah)$. As the characters of $L_2(\Ah)$ coincide with the characters of Ramond-twisted irreducible modules of $\cW_{-1}(E_6,f_\theta)$, we translate the module decompositions to those of $\cW_{-1}(E_6,f_\theta)$ in coset \eqref{cosE6} by simply replacing the effective $(\cB\mathrm{Vir}^{N=1}_{22,8})^{\rm e}$ to the original one denoted as $M_6$.  In the end, we obtain all 21 Ramond-twisted irreducible modules of $\cW_{-1}(E_6,f_\theta)$ and their decomposition to $L_2(A_5)\otimes M_6$ modules in coset construction \eqref{cosE6}. We summarize the results in the Table \ref{tab:WE6Ramondirr1} and \ref{tab:WE6Ramondirr2}. The character of $R_i$ is $\chi_{\mu_i}$ in \eqref{chiA0}. In Table \ref{tab:WE6Ramondirr1}, $R_6\cong \overline{R}_5$. In Table \ref{tab:WE6Ramondirr2}, $R_8\cong R_7\oplus \CC^{6}$ which can be deduced by comparing the character $\chi(R_8)-\chi(R_7)=6$. Besides, there exist seven more Ramond-twisted irreducible modules $\overline{R}_{7},\dots,\overline{R}_{13}$ which are complex conjugate to $R_7,\dots,R_{13}$. Each Ramond-twisted irreducible module $R_i$ can be regarded as an irreducible module $L_2(\Ah,\omega_i,\mu_i)$  in the sense of \cite{Lee:2024fxa}.

\begin{table}[ht]
\caption{All Ramond-twisted irreducible modules of $\cW_{-1}(E_6,f_\theta)$ with no character degeneracy in $\ZZ_3$ orbits. The conformal dimensions $h_{ij}$ are shown for each module $R_i=\bigoplus_j L_2(A_5,\lambda_j)\otimes M_6(h_{ij})$.}
\def\arraystretch{1.1}
    \centering
    \begin{tabular}{c|c|c|c|c|c|c|c|c|cccc} \hline
$\omega_i$ & $R_i$   &  $0$  &  $ 2w_3$    &   $ w_1+w_5$   & $w_2+w_4$   &   $ w_3$    &  $w_1+w_2$ & $w_4+w_5 $     & $\mu_i$
 \\ \hline    
$0$& $R_{0 }$ & $-3/22$   &   $  4/11  $   &   $  49/44  $   &    $  71/44  $   &     $ 73/352   $  &    $   997/352 $     &$   997/352 $     &    $0    $  \\
$ 2w_3$    & $R_{ 1}$ &     $ 3/2   $   &   $  0  $   &    $  11/4  $   &     $ 9/4   $  &    $   27/32 $   &    $ 143/32   $   & $ 143/32   $   &          $   18/11 $  \\ 
$ w_1+w_5$   &$R_{2 }$ &   $  15/22  $   &   $  13/11  $   &    $   -3/44 $   &     $  19/44  $  &    $ 9/352   $   &    $   229/352 $   & $   229/352 $   &     $  9/11  $     \\ 
$w_2+w_4$   &$R_{ 3}$ &    $  47/22  $   &   $  18/11  $   &    $ 17/44   $   &     $ -5/44   $  &    $  169/352  $   &    $  37/352  $  &    $  37/352  $   &     $  14/11  $    \\ 
 $ w_3$    &$R_{4 }$ &   $  1/2  $   &   $  1/11  $   &    $  37/44  $   &     $   15/44 $  &    $  -23/352  $   &    $  549/352  $ &    $  549/352  $   &     $ 8/11   $   \\ 
$w_1+w_2$ & $R_{ 5}$ &    $  65/22  $   &   $ 38/11   $   &    $ 9/44   $   &     $  31/44  $  &    $ 457/352   $&    $  -27/352   $   &    $  325/352  $   &     $  12/11  $      \\
$w_4+w_5 $ & $R_{ 6}$ &   $  65/22  $   &   $ 38/11   $   &    $ 9/44   $   &     $  31/44  $  &    $ 457/352   $   & $  325/352  $& $  -27/352  $   &    $  12/11  $     \\
\hline
     \end{tabular}
    \label{tab:WE6Ramondirr1}
\end{table}

\begin{table}[ht]
\caption{All Ramond-twisted irreducible modules of $\cW_{-1}(E_6,f_\theta)$ with character degeneracy two in $\ZZ_3$ orbits. The conformal dimensions $h_{ij}$ are shown for each module $R_i=\bigoplus_j L_2(A_5,\lambda_j)\otimes M_6(h_{ij})$.}
\def\arraystretch{1.1}
    \centering
    \begin{tabular}{c|c|c|c|c|c|c|c|c|ccccc} \hline
$\omega_i$ &  $R_i$   &  $w_5$  &  $ w_2+w_3$    &   $ w_2$   & $w_3+w_5$   &   $ 2w_1$    &  $2w_4$& $w_1+w_4  $     & $\mu_i$
 \\ \hline    
$w_{5}$&  $R_{7 }$ &    $  -27/352  $   &   $   325/352  $   &    $ 31/44   $   &     $  9/44  $  &    $  38/11  $   &    $   65/22 $   &  $  457/352  $   &    $  14/33  $    \\ 
$w_{2}+w_3$&  $R_{8 }$ &    $  325/352   $   &   $  -27/352  $   &    $  31/44  $   &     $  9/44  $  &    $  38/11  $   &    $  65/22  $   & $  457/352  $   &     $ 47/33   $    \\ 
$w_{2}$&  $R_{ 9}$ &    $ 37/352   $   &   $  37/352  $   &    $ -5/44  $   &     $   17/44 $  &    $   18/11 $   &    $   47/22 $   &   $  169/352  $   &   $   20/33 $    \\ 
$w_{3}+w_5$&  $R_{ 10}$ &    $  229/352  $   &   $  229/352  $   &    $  19/44  $   &     $  -3/44  $  &    $  13/11  $   &    $  15/22  $   &     $  9/352  $    &   $   38/33  $ \\ 
$2w_{1}$&  $R_{11 }$ &    $  143/32  $   &   $ 143/32   $   &    $ 9/4   $   &     $  11/4  $  &    $  0  $   &    $  3/2  $   &$   27/32 $   &     $  32/33  $    \\ 
$2w_{4}$&  $R_{ 12}$ &    $   997/352 $   &   $ 997/352   $   &    $  71/44  $   &     $   49/44 $  &    $   4/11 $   &    $   -3/22 $   &      $  73/352  $&  $  4/3  $    \\ 
$w_{1}+w_4$&  $R_{ 13}$ &    $  549/352  $   &   $  549/352  $   &    $  15/44  $   &     $  37/44  $  &    $ 1/11   $   &    $  1/2  $   &     $  -23/352  $  &   $  35/33 $  \\ 
\hline
     \end{tabular}
    \label{tab:WE6Ramondirr2}
\end{table}

\subsection{Irreducible modules of $\cW_{-1}(E_6,f_{\theta})$}
We now use a simple current of $L_2(A_5)$ to recover all irreducible modules of $\cW_{-1}(E_6,f_{\theta})$ from the Ramond-twisted ones. 
The simple current $L_2(A_5,2w_3,3/2)$ induces a $\ZZ_2$ automorphism for the 21 irreducible modules of $L_2(A_5)$. Besides the exchange between the vacuum and $L_2(A_5,2w_3,3/2)$, the simple current induces the following exchanges
\begin{align}\nonumber
 %L(A_5,0,0)\leftrightarrow   L(A_5,2w_3,3/2),\quad 
 L_2(A_5,2w_5,5/6)&\leftrightarrow L_2(A_5,2w_2,4/3),\\  \nonumber 
 L_2(A_5,2w_1,5/6)&\leftrightarrow L_2(A_5,2w_4,4/3),\\ \nonumber
 L_2(A_5,w_2,7/12 )&\leftrightarrow L_2(A_5,w_3+w_5,13/12 ),\\ \nonumber 
L_2(A_5,w_4,7/12 )&\leftrightarrow L_2(A_5,w_1+w_3, 13/12),\\ \nonumber
 L_2(A_5,w_5,35/96)&\leftrightarrow L_2(A_5,w_2+w_3,131/96),\\ \nonumber
 L_2(A_5,w_1,35/96)&\leftrightarrow L_2(A_5,w_3+w_4,131/96),\\ \nonumber
L_2(A_5,w_1+w_5,3/4 )&\leftrightarrow L_2(A_5,w_2+w_4, 5/4),\\ \nonumber L_2(A_5,w_4+w_5,33/32)&\leftrightarrow L_2(A_5,  w_1+w_2,33/32).
\end{align}
The rest three irreducible modules fusion to themselves
\begin{align}\nonumber
   L_2(A_5,w_3,21/32),\quad L_2(A_5,w_1+w_4,95/96),\quad  L_2(A_5,w_2+w_5,95/96).
\end{align}

By this $\ZZ_2$ automorphism for the $L_2(A_5)$ irreducible modules, we can easily obtain all 21 irreducible modules of $\cW_{-1}(E_6,f_{\theta})$ from the Ramond-twisted ones in Tables \ref{tab:WE6Ramondirr1} and \ref{tab:WE6Ramondirr2}. For example, the $R_1$ gives $N_0\cong \cW_{-1}(E_6,f_{\theta})$ in \eqref{isoE6}.

Finally we discuss the $\ZZ_2=\{1,\sigma\}$ automorphism group of $\cW_{-1}(E_6,f_{\theta})$. In \cite{Lee:2024fxa}, $w_1,w_3$ and $w_5$ of $A_5$ are called fermionic fundamental weights. A highest weight of $A_5$ irreducible is called bosonic/fermionic if it contains an even/odd number of fermionic fundamental weights. 
Then in Table \ref{tab:WE6Ramondirr1}, the first four columns from $0$ to $w_2+w_4$ are bosonic weights while  the last three columns from $w_3$ to $w_4+w_5$ are fermionic weights. According to the previous discussion, the even and odd parts of $\cW_{-1}(E_6,f_{\theta})$ are contributed by the bosonic and fermionic irreducible modules of of $L_2(A_5)$ respectively. This suggest that $\sigma(b)=b$ and $\sigma(f)=-f$ for any $b $ in $L_2(A_5)\otimes M_6$ modules in the first four columns and $f$ in $L_2(A_5)\otimes M_6$ modules in the last three columns in Table \ref{tab:WE6Ramondirr1}.

\subsection{Coset B}
$L_2(\Ah)$ is an intermediate vertex subalgebra of $L_2(E_6)$. 
We find the following coset construction holds: 
 \begin{align}\label{cosetBE6}
     \frac{L_2(E_6)}{ L_2(\Ah)}= \mathrm{Vir}^{(A_6,D_{12}),\rm eff}_{22,7}.
 \end{align}
The minimal model $\mathrm{Vir}_{22,7}$ has central charge $- {598}/{77}$ and effective central charge $ {74}/{77}$. It has $63$ irreducible modules. We compute their characters as $q$-series to sufficiently high degrees and then check the character relations for the coset \eqref{cosetBE6} for every irreducible module of $L_2(E_6)$. We find the coset gives precisely the type $(A_6,D_{12})$ non-diagonal modular invariant of effective $\mathrm{Vir}_{22,7}$. 
For the name of modular invariants of minimal models, we refer to the standard textbook \cite{CFT}. This non-diagonal modular invariant has 18 characters, among which three have degeneracy two. 
We write down this non-diagonal modular invariant by effective conformal dimensions:
\begin{align}
    Z=&\ \big|\chi^{\rm e}_{0}\big|^2+\big|\chi^{\rm e}_{\frac{3}{11}}\big|^2+\big|\chi^{\rm e}_{\frac{13}{11}}\big|^2+\big|\chi^{\rm e}_{\frac{30}{11}}\big|^2+\big|\chi^{\rm e}_{\frac{4}{11}}\big|^2+\big|\chi^{\rm e}_{\frac{15}{77}}\big|^2+\big|\chi^{\rm e}_{\frac{1}{77}}\big|^2+\big|\chi^{\rm e}_{\frac{435}{77}}\big|^2+\big|\chi^{\rm e}_{\frac{120}{77}}\big|^2\\ \nonumber
    &\,+\big|\chi^{\rm e}_{\frac{36}{77}}\big|^2+\big|\chi^{\rm e}_{\frac{6}{7}}\big|^2+\big|\chi^{\rm e}_{\frac{10}{77}}\big|^2+\big|\chi^{\rm e}_{\frac{3}{77}}\big|^2+\big|\chi^{\rm e}_{\frac{45}{77}}\big|^2+\big|\chi^{\rm e}_{\frac{171}{77}}\big|^2+2\big|\chi^{\rm e}_{\frac{15}{77}}\big|^2+2\big|\chi^{\rm e}_{\frac{136}{77}}\big|^2+2\big|\chi^{\rm e}_{\frac{54}{11}}\big|^2 .
\end{align}
This means that the extension of $\mathrm{Vir}_{22,7}$ has 21 irreducible modules.

We checked the character relations of coset construction \eqref{cosetBE6} for all irreducible modules of $L_2(E_6)$. The character relations can be upgraded to the module decompositions for $L_2(E_6)$ with carefully comparing with the module decompositions for $\Ah\subset E_6$ in \cite{Lee:2024fxa}. The information of the coset construction \eqref{cosetBE6} is collected in Table \ref{tab:CosetEA}, where the effective conformal dimensions $h_{ij}$ of the type $(A_6,D_{12})$ non-diagonal modular invariant of minimal model $\mathrm{Vir}_{22,7}$ in the decomposition are presented. The conformal dimension of $L_2(E_6,\omega_i)$ is $\mu_i$. The modules in the first tabular have no character degeneracy, while those in the second have complex conjugates. To be precise, in the second tabular in Table \ref{tab:CosetEA}, the $E_6$ modules $\bf 27,351,351'$ are displayed, which are decomposed into $\bf 6,\overline{21},\overline{21}'',120,\overline{210},\overline{210}',336$ modules of $\Ah$. 

\begin{table}[ht]
\caption{Coset construction $L_2(E_6,\omega_i,\mu_i)=\bigoplus_j L_2(\Ah,\lambda_j)\otimes \mathrm{Vir}^{(A_6,D_{12}),\rm eff}_{22,7}(h_{ij})$.}
\def\arraystretch{1.1}
    \centering
    \begin{tabular}{c|c|c|c|c|c|c|ccccccc} \hline
$\omega_i$     &  $0$  &  $ w_3 $    &   $ w_1+w_5 $   & $ w_1+w_2,w_4+w_5  $   &   $ w_2+w_4 $    &  $  2w_3 $    & $\mu_i$
 \\ \hline    
$0 $    &    $    0 $   &   $    3/11  $   &    $  13/11 $   &     $   54/11  $  &    $  30/11   $   &    $  4/11   $   &     $   0  $    \\
$ w_6$    &    $   6/7  $   &   $   10/77   $   &    $  3/77 $   &     $   136/77  $  &    $  45/77   $   &    $  171/77   $   &     $   6/7  $     \\
$w_1+w_5 $    &    $  23/7   $   &   $   120/77   $   &    $  36/77 $   &     $   15/77  $  &    $   1/77  $   &    $  435/77   $   &     $   9/7  $     \\
\hline
     \end{tabular}
     \\[+1mm]
      \begin{tabular}{c|c|c|c|c|c|c|ccccc} \hline
$\omega_i$     &  $w_1,w_3+w_4 $    &   $w_4 $   & $ 2w_5  $   &   $ w_2+w_5  $  &   $ w_1+w_3  $  &   $ 2w_2  $       & $\mu_i$
 \\ \hline    
$ w_1$    &    $  15/77   $   &   $  1/77    $   &    $  435/77 $   &     $  120/77   $  &    $  36/77   $   &    $  23/7   $   &   $ 13/21 $  \\
$w_4 $    &    $   136/77  $   &   $     45/77 $   &    $  171/77 $   &     $   10/77  $  &    $  3/77   $   &    $  6/7   $   &      $ 25/21 $  \\
$2w_5 $    &    $   54/11  $   &   $    30/11  $   &    $ 4/11  $   &     $   3/11  $  &    $  13/11   $   &    $  0   $   &        $ 4/3 $  \\
%$ $    &    $     $   &   $      $   &    $   $   &     $     $  &    $     $   &    $     $   &     $     $  &    $     $  & $  $  \\
\hline
     \end{tabular}
    \label{tab:CosetEA}
\end{table}

\section{Comments on $\cW_{-3}(E_8,f_{\theta})$ and $\cW_{-2}(E_8,f_\theta)$}\label{sec:E8high}
The $\cW$-algebras $\cW_{-h^\vee /6+n}(\mg,f_{\theta})$ with higher level for $1<n<h^\vee/6,n\in \ZZ$ are more difficult to study as they has more irreducible modules and it may be hard to find coset constructions like the $n=0,1$ cases. Nevertheless, assuming the Ramond-twisted modules of $\cW_{-h^\vee/6+n}(\mg,f_{\theta})$ are still effectively described by $L_{n+1}(\mg_I)$ -- the intermediate vertex subalgebra of $L_{n+1}(\mg)$, we are able to make some interesting observations for $\mg=E_8$ and $n=2,3$ cases, which may lead to explicit construction for these $\cW$-algebras in the future.

\subsection{Exceptional $\cW$-algebras and Hecke relations}
Exceptional $\cW$-algebras are $\cW_{k}(\mg,f)$ algebras with admissible level $k=-h^\vee +p/q$ and $(f,q)$ forming an exceptional pair \cite{kac2008rationality}. They are conjectured to be rational  \cite{kac2008rationality} and largely proved in \cite{arakawa2023rationality}. 
In \cite[Table 3]{arakawa2023rationality}, many exceptional subregular $\cW_{-h^\vee +p/q}(\mg,f_{\rm sreg})$ of ADE types are studied with various properties listed.  In particular, some isomorphisms with Virasoro minimal models or their extensions were proved. 

We observe that there may exist a series of Hecke relation between the exceptional subregular $\cW$-algebra with admissible levels and the minimal $\cW$-algebras with non-admissible levels for $\mg=E_{6,7,8}$:
\begin{align}
    \cW_{-h^\vee +(h^\vee +1)/(\frac56h^\vee +n)}(\mg,f_{\rm sreg}) \xrightarrow{\T_{\frac23h^\vee-1}}\cW_{-h^\vee/6+n}(\mg,f_{\theta}),\qquad n=0,1,\dots,h^\vee/6-1.
\end{align}
This suggests that the characters of all irreducible modules of the exceptional $\cW$-algebra are mapped to the characters of all Ramond-twisted irreducible modules of the minimal $\cW$-algebra by the Hecke operator. 
We checked this Hecke relation holds for $n=0,$ $\mg=E_6,E_7,E_8$ and $n=1,\mg=E_8$. For $n=1,\mg=E_7$, the Hecke relation holds if there exists isomorphism $\cW_{-18+\frac{19}{16}}(E_7,f_{\rm sreg} )\cong \cB\mathrm{Vir}^{N=1}_{16,2}$. Both sides of this conjectural isomorphism have central charge $-135/8$, effective central charge $9/8$ and $13$ irreducible modules. In particular, we compute the vacuum character of $\cB\mathrm{Vir}^{N=1}_{16,2}$ as 
$$
q^{\frac{45}{64}}(1+q^2+q^3+3 q^4+3 q^5+7 q^6+8 q^7+\dots).
$$
This is consistent with the fact that subregular $\cW$-algebras of type E are distinguished $\cW$-algebras, which have zero weight-one space \cite{arakawa2023rationality}.

In the following, we discuss the cases with $\mg=E_8$ and $n=2,3$. It is worthy to mention that Hecke operator induces Galois symmetry in modular tensor categories \cite{Harvey:2019qzs}. It would be interesting to further investigate the Hecke relations for $\cW$-algebras in MTCs.

\subsection{$\cW_{-3}(E_8,f_\theta)$}

The $\cW_{-3}(E_8,f_\theta)$ has central charge $c= {148}/{9}$, while $L_3(\Eh)$ has $c= {190}/{9}$. According to the study on $L_3(\Eh)$ in \cite{Lee:2023owa},  $\cW_{-3}(E_8,f_\theta)$ should have 12 irreducible modules. We want to use Hecke operator to produce the characters of $L_3(\Eh)$, which are expected to coincide with the characters of the 12 Ramond-twisted irreducible modules of $\cW_{-3}(E_8,f_\theta)$.

Inspired by the Hecke operator realizations for $L_1(\Eh)$ and $L_2(\Eh)$, we expect there exist a rational VOA $M_{{10}/{9}}$ with $c_{\rm eff}= {10}/{9}$, conductor $N=108$, and 12 irreducible modules and no character degeneracy such that 
\begin{align}
    L_5(G_2)&=\T_7  M_{{10}/{9}},\\
    L_3(\Eh)&=\T_{19}  M_{{10}/{9}}.
\end{align}
The affine VOA $L_5(G_2)$ has $c= {70}/{9}$ and 12 irreducible modules. We compute their conformal dimensions and exponents, and find 
the conductor $N=108$ and MLDE index $l=20$.  On the other hand, it was conjectured in \cite{Lee:2023owa} that $L_3(\Eh)$ has $c= {190}/{9}$ and also 12 irreducible modules and conductor $N=108$ and MLDE index $l=20$. In particular, all 12 conformal dimensions are given in 
\cite[Equation (4.10)]{Lee:2023owa}. Recall the basic property of Hecke operator is that it maps the (effective) conformal dimension by
$$
\T_p:h_i\xrightarrow{} p h_i\!\!\!\mod \ZZ
$$
We summarize the expected Hecke maps for the conformal dimensions in Table \ref{tab: Hecke1}.

\begin{table}[ht]
\caption{Map of conformal dimensions in the Hecke relations for $M_{10/9}$.}
\def\arraystretch{1.3}
    \centering
    \begin{tabular}{c|c|cccccccccccc} \hline
  $M_{10/9}$ & $h_i\!\!\mod \ZZ$  & $0$  & $\frac{1}{3}$  & $\frac{2}{3}$  & $\frac{2}{27}$  &  $\frac{11}{27}$  &      $\frac{20}{27}$  & $\frac{1}{9}$  & $\frac{2}{9}$  &  $\frac{4}{9}$&  $\frac{5}{9}$ &  $\frac{7}{9}$ & $\frac{8}{9}$ \\ 
 $\T_7$ & $L_5(G_2,h_i)$ &  $0$   &  $\frac{4}{3}$ &  $\frac{5}{3}$ & $\frac{14}{27}$ & $\frac{50}{27}$  & $\frac{32}{27}$    &  $\frac{7}{9}$ &  $\frac{14}{9}$ & $\frac{10}{9}$  &  $\frac{8}{9}$ &  $\frac{4}{9}$ &  $\frac{2}{9}$   \\ 
 $\T_{19}$ &  $L_3(\Eh,h_i)$  & $0$  &  $\frac{4}{3}$ & $\frac{5}{3}$  & $\frac{38}{27}$ &  $\frac{20}{27}$ & $\frac{56}{27}$    &  $\frac{19}{9}$  &  $\frac{20}{9}$ &  $\frac{22}{9}$ & $\frac{14}{9}$ &  $\frac{16}{9}$ &  $\frac{ 8}{9}$   \\ \hline    
     \end{tabular}
    \label{tab: Hecke1}
\end{table}

We further conjecture that the $M_{10/9}$ is the effective description of the exceptional $\cW$-algebra $\cW_{-30+\frac{31}{27}}(E_8,f_{\rm sreg}) $ studied in \cite{arakawa2023rationality}. This rational $\cW$-algebra  has central charge $-590/9$ and effective central charge $10/9$ and $12$ irreducible modules, which satisfy the necessary conditions for $M_{10/9}$.  
The vacuum of $\cW_{-30+\frac{31}{27}}(E_8,f_{\rm sreg}) $ has effective conformal dimension $25/9 $, which corresponds to the item $7/9$ in the first row of Table \ref{tab: Hecke1}. This conjecture suggests surprising  connections among $\cW_{-30+\frac{31}{27}}(E_8,f_{\rm sreg}) $, $L_5(G_2)$ and $\cW_{-3}(E_8,f_\theta)$ in MTCs.

\subsection{$\cW_{-2}(E_8,f_\theta)$}
The $\cW_{-2}(E_8,f_\theta)$ has central charge $c= { 142}/{7}$, while $L_4(\Eh)$ has $c={190}/{7}$. The study on $L_4(\Eh)$ in \cite{Lee:2023owa} suggests that $\cW_{-2}(E_8,f_\theta)$ should have 25 irreducible modules. Again we would like to use Hecke operator to produce the characters of $L_4(\Eh)$, which are expected to coincide with the characters of the Ramond-twisted irreducible modules of $\cW_{-2}(E_8,f_\theta)$.

In this case, we expect there exist a rational VOA $M_{{10}/{7}}$ with $c_{\rm eff}= {10}/{7}$, conductor $N=84$, and 25 irreducible modules and no character degeneracy such that
\begin{align}
    \T_{13}M_{10/7}&=L_5(F_4),\\
    \T_{19}M_{10/7}&=L_4(\Eh).
\end{align} 
The affine VOA $L_5(F_4)$ has 25 irreducible modules. We compute their conformal dimensions and exponents and find 
the conductor $N=84$ and the MLDE index $l=160$. It was predicted in \cite{Lee:2023owa}  that the intermediate VOA $L_4(\Eh)$ also has 25 irreducible modules and conductor $N=84$ and the MLDE index $l=160$. In particular, all conformal dimensions are predicted in \cite[Equation (4.10)]{Lee:2023owa}. We summarize the expected Hecke maps for the conformal dimensions in Table \ref{tab: Hecke2}.

\begin{table}[ht]
\caption{Map of conformal dimensions in the Hecke relations for $M_{10/7}$.}
\def\arraystretch{1.3}
    \centering
    \begin{tabular}{c|c|cccccccccccc} \hline
 \! $ \!\! \!M_{10/7}  \!\!$ \! & $h_i\!\!\mod \ZZ$  & $0_3\!$  & $(\frac{1}{2})_2$  &  $\frac{1}{4}$ & $  (\frac{1}{14})_2 $  & $ (\frac17 )_3 $  & $ \frac{2}{7}  $  & $  \frac{3}{7} $& $ (\frac{4}{7} )_3 $& $ \frac{5}{7}  $& $ (\frac{6}{7})_3  $  & $ \frac{3}{28}  $  & $ \frac{23}{28}  \!\!$  \\ 
 $\T_{13}$ & $L_5(F_4,h_i)$ &  $0,2,2\!$   &  $ \frac32,\frac52$  &  $\frac{9}{4}$ & $ \frac{13}{14},\frac{27}{14}  $  & $ \frac{6}{7},\frac{13}{7} ,\frac{20}{7} $  & $ \frac{12}{7}  $  & $ \frac{ 18}{7}  $ & $ \frac{3}{7},\frac{10}{7},\frac{17}{7}  $& $ \frac97  $& $ \frac{8}{7},\frac{15}{7},\frac{15}{7}  $& $ \frac{39}{28}  $  & $ \frac{75}{28} \!\! $ \\ 
 $\T_{19}$ &  $\! L_4(\Eh,h_i)\! $ & $0,2,3\!$  &  $ \frac32,\frac52$ &  $\frac{11}{4}$  & $  \frac{19}{14},\frac{33}{14} $  & $ \frac{5}{7},\frac{12}{7} ,\frac{19}{7} $  & $\frac{17}{7}    $  & $  \frac{15}{7} $ & $ \frac67,\frac{20}{7},\frac{20}{7}  $& $ \frac{18}{7}  $& $ \frac{9}{7},\frac{16}{7},\frac{23}{7}  $ & $  \frac{57}{28} $  & $ \frac{45}{28} \!\! $  \\ \hline    
     \end{tabular}
   % \caption{ }
    \label{tab: Hecke2}
\end{table}

We further conjecture that the $M_{10/7}$ is the effective description of the exceptional $\cW$-algebra $\cW_{-30+\frac{31}{28}}(E_8,f_{\rm sreg}) $ studied in \cite{arakawa2023rationality}. This rational $\cW$-algebra with admissible level has central charge $-830/7$, effective central charge $10/7$ and $25$ irreducible modules, which satisfy the necessary conditions for $M_{10/7}$. The vacuum of $\cW_{-30+\frac{31}{28}}(E_8,f_{\rm sreg}) $ has effective conformal dimension $5$, which corresponds to one conformal dimension of the item $0_3$ in the first row of Table \ref{tab: Hecke2}. This conjecture suggests surprising connections among $\cW_{-30+\frac{31}{28}}(E_8,f_{\rm sreg}) $, $L_5(F_4)$ and $\cW_{-2}(E_8,f_\theta)$ in MTCs.

\bigskip

\noindent
\textbf{Acknowledgements}
The author would like to thank T. Arakawa, T. Creutzig, C. Dong and K. Kawasetsu for inspiring discussions. The author is also grateful to J. van Ekeren for sharing relevant data of exceptional subregular $\cW$-algebras. He is supported in part by the Swedish Olle Engkvists Stiftelse Grant No.2180108. He thanks the Galileo Galilei Institute for Theoretical Physics for the hospitality during the completion of this work.

\bibliographystyle{plainnat}

\begin{thebibliography}{34}
\providecommand{\natexlab}[1]{#1}
\providecommand{\url}[1]{\texttt{#1}}
\expandafter\ifx\csname urlstyle\endcsname\relax
  \providecommand{\doi}[1]{doi: #1}\else
  \providecommand{\doi}{doi: \begingroup \urlstyle{rm}\Url}\fi

\bibitem[Arakawa(2015{\natexlab{a}})]{arakawa2015associated}
Tomoyuki Arakawa.
\newblock Associated varieties of modules over {K}ac--{M}oody algebras and {$C_2$}-cofiniteness of {W}-algebras.
\newblock \emph{International Mathematics Research Notices}, 2015\penalty0 (22):\penalty0 11605--11666, 2015{\natexlab{a}}.

\bibitem[Arakawa(2015{\natexlab{b}})]{arakawa2015rationality}
Tomoyuki Arakawa.
\newblock Rationality of {W}-algebras: principal nilpotent cases.
\newblock \emph{Annals of Mathematics}, pages 565--604, 2015{\natexlab{b}}.

\bibitem[Arakawa and Moreau(2018)]{AM}
Tomoyuki Arakawa and Anne Moreau.
\newblock Joseph ideals and lisse minimal {$W$}-algebras.
\newblock \emph{J. Inst. Math. Jussieu}, 17\penalty0 (2):\penalty0 397--417, 2018.

\bibitem[Arakawa and van Ekeren(2023)]{arakawa2023rationality}
Tomoyuki Arakawa and Jethro van Ekeren.
\newblock Rationality and fusion rules of exceptional $ \mathcal{W}$-algebras.
\newblock \emph{Journal of the European Mathematical Society}, 25\penalty0 (7):\penalty0 2763--2813, 2023.

\bibitem[Bershadsky et~al.(1985)Bershadsky, Knizhnik, and Teitelman]{Bershadsky:1985dq}
M.~A. Bershadsky, V.~G. Knizhnik, and M.~G. Teitelman.
\newblock {Superconformal Symmetry in Two-Dimensions}.
\newblock \emph{Phys. Lett. B}, 151:\penalty0 31--36, 1985.

\bibitem[Creutzig and Linshaw(2022)]{Creutzig:2021dda}
Thomas Creutzig and Andrew~R. Linshaw.
\newblock {Trialities of orthosymplectic W-algebras}.
\newblock \emph{Adv. Math.}, 409:\penalty0 108678, 2022.

\bibitem[Cvitanovic(2008)]{Cvitanovic:2008zz}
Predrag Cvitanovic.
\newblock \emph{{Group theory: Birdtracks, Lie's and exceptional groups}}.
\newblock Pr. Princeton, 2008.

\bibitem[Deligne(1996)]{Deligne}
Pierre Deligne.
\newblock La s\'{e}rie exceptionnelle de groupes de {L}ie.
\newblock \emph{C. R. Acad. Sci. Paris S\'{e}r. I Math.}, 322\penalty0 (4):\penalty0 321--326, 1996.

\bibitem[Di~Francesco et~al.(1997)Di~Francesco, Mathieu, and S\'{e}n\'{e}chal]{CFT}
Philippe Di~Francesco, Pierre Mathieu, and David S\'{e}n\'{e}chal.
\newblock \emph{Conformal field theory}.
\newblock Graduate Texts in Contemporary Physics. Springer-Verlag, New York, 1997.

\bibitem[Duan et~al.(2022)Duan, Lee, and Sun]{Duan:2022ltz}
Zhihao Duan, Kimyeong Lee, and Kaiwen Sun.
\newblock {Hecke relations, cosets and the classification of 2d RCFTs}.
\newblock \emph{JHEP}, 09:\penalty0 202, 2022.

\bibitem[Duan et~al.(2023)Duan, Lee, Lee, and Li]{Duan:2022kxr}
Zhihao Duan, Kimyeong Lee, Sungjay Lee, and Linfeng Li.
\newblock {On classification of fermionic rational conformal field theories}.
\newblock \emph{JHEP}, 02:\penalty0 079, 2023.

\bibitem[Fasquel(2022)]{fasquel2022geometry}
Justine Fasquel.
\newblock \emph{Geometry and new rational W-algebras}.
\newblock PhD thesis, Universit{\'e} de Lille, 2022.

\bibitem[Feigin and Frenkel(1990)]{Feigin:1990pn}
B.~Feigin and E.~Frenkel.
\newblock {Quantization of the Drinfeld-Sokolov reduction}.
\newblock \emph{Phys. Lett. B}, 246:\penalty0 75--81, 1990.

\bibitem[Frenkel et~al.(1992)Frenkel, Kac, and Wakimoto]{Frenkel:1992ju}
Edward Frenkel, Victor Kac, and Minoru Wakimoto.
\newblock {Characters and fusion rules for W algebras via quantized Drinfeld-Sokolov reductions}.
\newblock \emph{Commun. Math. Phys.}, 147:\penalty0 295--328, 1992.

\bibitem[Friedan et~al.(1984)Friedan, Qiu, and Shenker]{Friedan:1983xq}
Daniel Friedan, Zong-an Qiu, and Stephen~H. Shenker.
\newblock {Conformal Invariance, Unitarity and Two-Dimensional Critical Exponents}.
\newblock \emph{Phys. Rev. Lett.}, 52:\penalty0 1575--1578, 1984.

\bibitem[Friedan et~al.(1985)Friedan, Qiu, and Shenker]{Friedan:1984rv}
Daniel Friedan, Zong-an Qiu, and Stephen~H. Shenker.
\newblock {Superconformal Invariance in Two-Dimensions and the Tricritical Ising Model}.
\newblock \emph{Phys. Lett. B}, 151:\penalty0 37--43, 1985.

\bibitem[Goddard et~al.(1986)Goddard, Kent, and Olive]{Goddard:1986ee}
P.~Goddard, A.~Kent, and David~I. Olive.
\newblock {Unitary Representations of the Virasoro and Supervirasoro Algebras}.
\newblock \emph{Commun. Math. Phys.}, 103:\penalty0 105--119, 1986.

\bibitem[Harvey and Wu(2018)]{Harvey:2018rdc}
Jeffrey~A. Harvey and Yuxiao Wu.
\newblock {Hecke Relations in Rational Conformal Field Theory}.
\newblock \emph{JHEP}, 09:\penalty0 032, 2018.

\bibitem[Harvey et~al.(2020)Harvey, Hu, and Wu]{Harvey:2019qzs}
Jeffrey~A. Harvey, Yichen Hu, and Yuxiao Wu.
\newblock {Galois Symmetry Induced by Hecke Relations in Rational Conformal Field Theory and Associated Modular Tensor Categories}.
\newblock \emph{J. Phys. A}, 53\penalty0 (33):\penalty0 334003, 2020.

\bibitem[Kac and Wakimoto(2004)]{Kac:2003jh}
Victor~G. Kac and Minoru Wakimoto.
\newblock {Quantum reduction and representation theory of superconformal algebras}.
\newblock \emph{Adv. Math.}, 185\penalty0 (2):\penalty0 400--458, 2004.

\bibitem[Kac and Wakimoto(2008)]{kac2008rationality}
Victor~G. Kac and Minoru Wakimoto.
\newblock On rationality of {W}-algebras.
\newblock \emph{Transformation Groups}, 13\penalty0 (3):\penalty0 671--713, 2008.

\bibitem[Kac et~al.(2024)Kac, Frajria, and Papi]{Kac:2024kvv}
Victor~G. Kac, Pierluigi~M\"oseneder Frajria, and Paolo Papi.
\newblock {Unitarity of minimal $W$-algebras and their representations II: Ramond sector}.
\newblock Preprint, 2024.
\newblock URL \url{https://arxiv.org/abs/2405.19090}.


\bibitem[Kawasetsu(2014)]{Kawasetsu}
Kazuya Kawasetsu.
\newblock The intermediate vertex subalgebras of the lattice vertex operator algebras.
\newblock \emph{Lett. Math. Phys.}, 104\penalty0 (2):\penalty0 157--178, 2014.

\bibitem[Kawasetsu(2018)]{Kawasetsu:2018irs}
Kazuya Kawasetsu.
\newblock {${\mathcal{W}}$-algebras with Non-admissible Levels and the Deligne Exceptional Series}.
\newblock \emph{Int. Math. Res. Not.}, 2018\penalty0 (3):\penalty0 641--676, 2018.

\bibitem[Kawasetsu and Sakai(2018)]{Kawasetsu:2018tzs}
Kazuya Kawasetsu and Yuichi Sakai.
\newblock {Modular linear differential equations of fourth order and minimal W -algebras}.
\newblock \emph{J. Algebra}, 506:\penalty0 445--488, 2018.

\bibitem[Landsberg and Manivel(2004)]{LMseries}
J.~M. Landsberg and L.~Manivel.
\newblock Series of {L}ie groups.
\newblock \emph{Michigan Math. J.}, 52\penalty0 (2):\penalty0 453--479, 2004.

\bibitem[Landsberg and Manivel(2006)]{LM}
J.~M. Landsberg and L.~Manivel.
\newblock The sextonions and {$E_{7\frac 12}$}.
\newblock \emph{Adv. Math.}, 201\penalty0 (1):\penalty0 143--179, 2006.

\bibitem[Lee and Sun(2023)]{Lee:2022yic}
Kimyeong Lee and Kaiwen Sun.
\newblock {Hecke relations among 2d fermionic RCFTs}.
\newblock \emph{JHEP}, 09:\penalty0 044, 2023.

\bibitem[Lee et~al.(2024{\natexlab{a}})Lee, Sun, and Wang]{Lee:2023owa}
Kimyeong Lee, Kaiwen Sun, and Haowu Wang.
\newblock {On intermediate Lie algebra $E_{7+1/2}$}.
\newblock \emph{Lett. Math. Phys.}, 114\penalty0 (1):\penalty0 13, 2024{\natexlab{a}}.

\bibitem[Lee et~al.(2024{\natexlab{b}})Lee, Sun, and Wang]{Lee:2024fxa}
Kimyeong Lee, Kaiwen Sun, and Haowu Wang.
\newblock {On Intermediate Exceptional Series}.
\newblock Preprint, 2024.
\newblock URL \url{https://arxiv.org/abs/2403.14311}.

\bibitem[Mathur et~al.(1988)Mathur, Mukhi, and Sen]{Mathur:1988na}
Samir~D. Mathur, Sunil Mukhi, and Ashoke Sen.
\newblock {On the Classification of Rational Conformal Field Theories}.
\newblock \emph{Phys. Lett. B}, 213:\penalty0 303--308, 1988.

\bibitem[Milas et~al.(2023)Milas, Penn, and Sadowski]{Milas:2023cwx}
Antun Milas, Michael Penn, and Christopher Sadowski.
\newblock {$S_3$-permutation orbifolds of Virasoro vertex algebras}.
\newblock \emph{J. Pure Appl. Algebra}, 227:\penalty0 107378, 2023.

\bibitem[Vogel(1999)]{vogel1999universal}
Pierre Vogel.
\newblock The {U}niversal {L}ie algebra.
\newblock Preprint, 1999.
\newblock URL \url{https://webusers.imj-prg.fr/~pierre.vogel/grenoble-99b.pdf}.

\bibitem[Zamolodchikov(1985)]{zamolodchikov1985infinite}
Aleksandr~B. Zamolodchikov.
\newblock Infinite additional symmetries in two-dimensional conformal quantum field theory.
\newblock \emph{Teoreticheskaya i Matematicheskaya Fizika}, 65\penalty0 (3):\penalty0 347--359, 1985.

\end{thebibliography}
\bibliofont

\end{document}